\documentclass[apj]{emulateapj}
\usepackage{amssymb,latexsym,amsmath}
\usepackage{apjfonts}
\shorttitle{Growth of Black Holes and Bulges in CDGs}
\shortauthors{Rafferty et al.}
\slugcomment{Accepted to The Astrophysical Journal}

\begin{document}

\title{The Feedback-Regulated Growth of Black Holes and Bulges through Gas Accretion and Starbursts in Cluster Central Dominant Galaxies}
       
\author{D. A. Rafferty and B. R. McNamara}
\affil{Department of Physics and Astronomy, Ohio University, Athens, OH 45701}

\author{P. E. J. Nulsen\altaffilmark{1}}
\affil{Harvard-Smithsonian Center for Astrophysics, 60 Garden St., Cambridge,
MA 02138 }
\altaffiltext{1}{On leave from the School of Engineering Physics, University of Wollongong, Wollongong, NSW 2522, Australia}

\and

\author{M. W. Wise}
\affil{Astronomical Institute ``Anton Pannekoek,'' University of Amsterdam, Kruislaan 403, 1098 SJ Amsterdam, The Netherlands}

\begin{abstract}
We present an analysis of the growth of black holes through accretion and bulges through star formation in 33 galaxies at the centers of cooling flows. Most of these systems show evidence of cavities in the intracluster medium (ICM) inflated by radio jets emanating from their active galactic nuclei (AGN). We present a new and extensive analysis of X-ray cavities in these systems. We find that AGN are  energetically able to balance radiative losses (cooling) from the ICM in more than half of our sample. Using a subsample of 17 systems, we examine the relationship between cooling and star formation. We find that the star formation rates are approaching or are comparable to X-ray and far UV limits on the rates of gas condensation onto the central galaxy. The remaining radiative losses could be offset by AGN feedback. The vast gulf between radiative losses and the sink of cooling material, which has been the primary objection to cooling flows, has narrowed and, in some cases, is no longer a serious issue. Using the cavity (jet) powers, we place strong lower limits on the rate of growth of supermassive black holes in central galaxies, and we find that they are growing at an average rate of $\sim 0.1$ M$_{\sun}$ yr$^{-1},$ with some systems growing as quickly as $\sim 1$ M$_{\sun}$ yr$^{-1}$. We find a trend between bulge growth (star formation) and black hole growth that is approximately in accordance with the slope of the local (Magorrian) relation between black hole and bulge mass. However, the large scatter in the trend suggests that bulges and black holes do not always grow in lock step. With the exception of the rapidly accreting supercavity systems (e.g, MS 0735.6+7421), the black holes are accreting well below their Eddington rates. Most systems could be powered by Bondi accretion from the hot ICM, provided the central gas density increases into the Bondi radius as $\rho \propto r^{-1}$. However, if the slope of the gas density profile flattens into a core, as observed in M87, Bondi accretion is unlikely to be driving the most powerful outbursts.
\end{abstract}


\section{Introduction}
The intracluster gas at the center of a majority of galaxy clusters has a cooling time less than $10^{10}$ yr \citep{edge92,pere98}. In the absence of a source of heat, this gas should cool, resulting in a slow inward flow of material know as a ``cooling flow'' \citep{fabi94}. While early observations of this gas made in X-rays supported this picture, observations of the clusters at other wavelengths did not.  Optical data implied star formation rates in the central galaxy of only a few percent of the derived cooling rates \citep[e.g.,][]{mcna89}, and radio observations found less cold gas than predicted \citep[e.g.,][]{edge01}.  However, recent high-resolution X-ray spectra from XMM-\textit{Newton} do not show the features expected if large amounts of gas are cooling below $kT \sim 2$ keV \citep{pete03,kaast04}. In addition, high spatial resolution \textit{Chandra} X-ray Observatory images reveal that AGN can have a large heating effect (via jets and winds) on the ICM, supplying enough heat in some systems to offset radiation losses \citep{birz04}. The emerging picture of cooling flows is one in which most (but not all) of the cooling is roughly balanced by heating from AGN feedback, resulting in a moderate cooling flow \citep[e.g,][]{pedl90,soke01,binn05,soke06}.

In this regulated-cooling scenario, net cooling from the ICM would lead to condensation of gas onto the central galaxy, driving the star formation observed in many systems \citep[e.g.,][]{john87,mcna89}. If this scenario is correct, the star formation rates should on average be comparable to the rate of gas observed to be condensing out of the ICM. Studies of a small number of systems with reliable star formation and cooling rates have shown that the rates are converging, and in some cases are in rough agreement \citep[e.g.,][]{mcna03,mcna04,mcna06}. The quality and quantity of data from \textit{Chandra} and XMM-\textit{Newton} now makes it possible to construct a significantly larger sample of such systems to better understand the possible connection between star formation and net cooling in cooling flow clusters. 

Additionally, the high-resolution data from \textit{Chandra} are useful in studies of the nature of the feedback mechanism that may be preventing large amounts of intracluster gas from cooling. Chandra images of galaxy clusters have revealed many large-scale interactions between the ICM and the central AGN, the best-known examples of which are the Perseus cluster \citep{boeh93,fabi00,schm02,fabi02b,fabi03a,fabi03b,fabi06}, Abell 2052 \citep{blan01,blan03}, and Hydra~A \citep{mcna00,davi01,nuls02,nuls05b}. In these systems, the radio jets of the AGN have pushed out cavities in the cluster's atmosphere, creating surface-brightness depressions in X-ray images that are correlated with the lobes' radio emission, such that the radio emission fills the depression in X-rays. The lower emissivities of the depressions imply that they are low-density cavities in the ICM, and therefore should rise buoyantly in the cluster's atmosphere \citep{chur01}.  By measuring the surrounding pressure and volume of the cavities using the X-ray data, one can derive the work done by the radio source on the ICM in inflating the cavities, giving a direct measurement of the non-radiative energy released during the outburst. Measurements of this energy, combined with measurements of the star formation and cooling rates, can be used to investigate possible feedback scenarios that may govern the growth of the central dominant galaxy (CDG, as distinct from the more strictly defined cD) and its central supermassive black hole.

Such feedback has implications for the more general problem of galaxy formation. The large-scale distribution of mass in the Universe is well modeled by the standard hierarchical cold dark matter (CDM) cosmology \citep{whit78}. In this model, larger dark matter halos form through the merging of smaller ones, while their gravitationally bound baryons cool and condense into the progenitors of the galaxies we see today \citep{cole91,blan92}. This picture successfully explains much of the observed matter distribution. However, a persistent problem in simulations that include only gravitational heating is a failure to reproduce the truncation of the high-luminosity end of the galaxy luminosity function \citep{bens03}. This problem stems from excessive cooling of baryons in the cores of halos, resulting in a population of massive galaxies, far larger than even the enormous cD galaxies at the cores of clusters \citep{sija06}. Instead of residing in the cD galaxy as predicted by simulations, most baryons are found in the hot ICM. 

This problem may have a solution in non-gravitational heating by supernovae and AGN. Supernovae are essential for enriching the ICM to observed levels \citep{metz94, borg02} and may play a significant heating role in smaller galaxies, but their feedback energies are too small and localized to truncate cooling in massive galaxies \citep{borg02}. Furthermore, in the closely related preheating problem, they have difficulty supplying enough heat to boost the entropy level of the hot gas to the observed levels \citep{wu00,borg05}. Energetically, AGN heating appears to be the most likely mechanism to severely reduce the supply of gas from the hot ICM in galaxies above a certain size and to explain observed entropy profiles \citep{bens03,scan04,dona06,voitd05,voit05}.

Current theory posits that AGN are powered by the accretion of material onto a central black hole.  Gravitational binding energy of the accreting material powers radiation and outflows from AGN as the black hole grows. The relativistic jets that are revealed by their synchrotron emission are a product of this process. The remaining accreting material goes to increasing the mass of the black hole. In a sense, AGN are the ``smoking guns'' of black hole growth. The fraction of accreted power that re-emerges from an AGN and its partitioning between radiation and outflows is not well understood, but probably depends on accretion rate \citep[e.g.,][]{rees82,nara94,abra95,chur05}. We can place lower limits on the AGN's power using estimates of the power required to create the cavities associated with the radio lobes. This power may then be used to infer the minimum growth rate of the black hole. 

As presented by \citet{ferr00} and \citet{gebh00}, a correlation exists between the mass of the central black hole ($M_{\rm{BH}}$) and the velocity dispersion ($\sigma$) of the galaxy's bulge. This correlation suggests that the large-scale properties of the galaxy and the small-scale properties of the black hole are related \citep[the ``Magorrian relation'',][]{mago98}. Estimates of the current growth rates of the black hole may be compared to the large-scale properties of the galaxy (such as the star formation rate) to trace the present day impact of bulge and black hole growth on this connection.

In this paper, we use star formation rates, ICM cooling rates, and AGN heating rates for a sample of cooling flows to investigate the relationships between star formation and cooling and between the growth rates of black holes and their host galaxies. We assume $H_0 = 70$~km s$^{-1}$ Mpc$^{-1}$, $\Omega_{\Lambda} = 0.7$, and $\Omega_{\rm{M}} = 0.3$ throughout.

\begin{deluxetable*}{llcccccl}
\tabletypesize{\scriptsize}
\tablewidth{0pt}
\tablecaption{Sample Properties. \label{T:sample}}
\tablehead{
	\colhead{} & \colhead{} & \colhead{} & \colhead{$\sigma_{\rm{c}}$\tablenotemark{a}} & \colhead{} & \colhead{} & \colhead{$M_{\rm{bulge}}$\tablenotemark{c}} & \colhead{} \\
	\colhead{System\phn} & \colhead{$z$} & \colhead{CDG} & \colhead{(km s$^{-1}$)} & \colhead{$M_{K}$\tablenotemark{b}} & \colhead{$M_{R}$\tablenotemark{b}} & \colhead{(10$^{11}$ $M_{\sun}$)} & \colhead{References} }
\startdata
	A85	    & 0.055  & PGC 002501		& 340$\pm$9		& $-26.72 \pm 0.04$ & $-24.80 \pm 0.08$	& 31$\pm$1 & 10, 26	\\
	A133        & 0.060  & ESO 541-013             & \nodata         	& $-26.36 \pm 0.06$ & $-24.18 \pm 0.05$	& 17.9$\pm$0.4 & 18, 19	\\
	A262        & 0.016  & NGC 708                 & 255$\pm8$      	& $-25.65 \pm 0.03$ & $-22.77 \pm 0.02$	& 4.9$\pm$0.1  & 5	\\
	Perseus     & 0.018  & NGC 1275                & 247$\pm$10  		& $-26.23 \pm 0.04$ & $-24.25 \pm 0.01$	& 19.2$\pm$0.1 & 12, 13, 15, 43\\ 
   	2A 0335+096 & 0.035  & PGC 013424              & \nodata         	& $-26.15 \pm 0.05$ & $-24.18 \pm 0.13$	& 18$\pm$1 & 31	\\
	A478        & 0.081  & PGC 014685              & \nodata         	& $-26.64 \pm 0.07$ & $-24.66 \pm 0.10$	& 28$\pm$1 & 44, 48	\\ 
	MS 0735.6+7421& 0.216  & PGC 2760958 	       & \nodata		& $-26.37 \pm 0.17$ & $-24.51 \pm 0.10$	& 24$\pm$1 & 36	\\
	PKS 0745-191& 0.103  & PGC 021813              & \nodata		& $-26.82 \pm 0.09$ & $-24.63 \pm 0.10$	& 27$\pm$1 & 21	\\
	4C 55.16    & 0.242  & PGC 2506893 		& \nodata		& $-26.10 \pm 0.13$ & $-24.75 \pm 0.50$	& 30$\pm$8 & 22	\\
	Hydra A     & 0.055  & PGC 026269              & 322$\pm$20 		& $-25.91 \pm 0.06$ & $-24.67 \pm 0.05$	& 28.2$\pm$0.7 & 9, 33, 38, 40	\\
	RBS 797     & 0.350  & \nodata                 & \nodata         	& \nodata	    & \nodata		& \nodata  & 45	\\  
	Zw 2701	    & 0.214  & PGC 2401970		& \nodata		& $-26.26 \pm 0.17$ & $-24.75 \pm 0.10$	& 30$\pm$1 & 2	\\
	Zw 3146	    & 0.291  & 2MASX J10233960+0411116 & \nodata		& $-27.67 \pm 0.14$ & \nodata		& \nodata  & 2, 23	\\
	A1068\tablenotemark{d}& 0.138  & PGC 093944	       & \nodata		& $-26.71 \pm 0.08$ & $-25.07 \pm 0.21$	& 41$\pm$4 & 35, 50	\\
	M84         & 0.0035 & M84                     & 298$\pm$2  		& $-24.69 \pm 0.02$ & $-22.62$		& 4.3 	   & 16	\\
	M87         & 0.0042 & M87                     & 341$\pm$3    		& $-25.55 \pm 0.02$ & $-23.61$		& 11	   & 16, 51	\\
	Centaurus   & 0.011  & NGC 4696                & 257$\pm$6 	  	& $-26.02 \pm 0.02$ & $-23.70 \pm 0.01$	& 11.6$\pm$0.1  & 42, 45	\\	
	HCG 62      & 0.014  & NGC 4778                & \nodata         	& $-25.26 \pm 0.03$ & \nodata		& \nodata  & 49	\\
	A1795       & 0.063  & PGC 049005              & 294$\pm$10   		& $-26.50 \pm 0.08$ & $-23.86 \pm 0.10$	& 13.4$\pm$0.6 & 11	\\ 
	A1835	    & 0.253  & 2MASX J14010204+0252423 & \nodata		& $-27.36 \pm 0.14$ & \nodata		& \nodata  & 37, 46\\
	PKS 1404-267& 0.022  & IC 4374			& 258$\pm$7		& $-25.30 \pm 0.03$ & $-22.93 \pm 0.20$	& 5.7$\pm$0.5 & 25 \\
	MACS J1423.8+2404& 0.545  & \nodata			& \nodata		& \nodata	    & \nodata		& \nodata  & 1	\\
	A2029       & 0.077  & PGC 054167              & 366$\pm$9 		& $-27.44 \pm 0.05$ & $-24.39 \pm 0.02$	& 21.9$\pm$0.2 & 7	\\
	A2052       & 0.035  & UGC 09799               & 259$\pm$11  		& $-26.27 \pm 0.06$ & $-23.62$		& 11 	   & 3, 4\\
	MKW 3S      & 0.045  & NGC 5920                & \nodata         	& $-25.55 \pm 0.06$ & $-23.67 \pm 0.05$	& 11.2$\pm$0.3  & 30, 32	\\
	A2199       & 0.030  & NGC 6166                & 302$\pm$4   		& $-26.37 \pm 0.03$ & $-24.03 \pm 0.03$	& 15.7$\pm$0.2 & 24	\\
	Hercules A  & 0.154  & PGC 059117		& \nodata		& $-26.45 \pm 0.11$ & $-23.95 \pm 0.50$	& 15$\pm$4 & 39	\\
	3C 388	    & 0.092  & PGC 062332		& 365$\pm$23		& $-26.24 \pm 0.06$ & $-24.46 \pm 0.50$	& 23$\pm$6 & 28, 29	\\
	3C 401	    & 0.201  & PGC 2605547 		& \nodata		& \nodata  	    & $-23.43 \pm 0.50$	& 9$\pm$2  & 41	\\
	Cygnus A    & 0.056  & PGC 063932              & \nodata       		& $-26.70 \pm 0.06$ & $-23.47 \pm 0.35$	& 9$\pm$2  & 27, 47	\\
	Sersic 159/03& 0.058 & ESO 291-009 		& \nodata	& $-26.26 \pm 0.10$ & $-23.68 \pm 0.39$	& 11$\pm$2  & 52	\\
	A2597       & 0.085  & PGC 071390              & 222$\pm$18  		& $-25.55 \pm 0.11$ & $-23.49 \pm 0.21$	& 9$\pm$1  & 8, 34	\\
	A4059       & 0.048  & ESO 349-010             & 296$\pm$49   		& $-26.74 \pm 0.05$ & $-25.00 \pm 0.02$	& 38.2$\pm$0.4 & 5, 19	
\enddata
\tablerefs{(1) Allen et al.\ 2004; (2) Bauer et al.\ 2005; (3) Blanton et al.\ 2001; (4) Blanton et al.\ 2003; (5) Blanton et al.\ 2004; (6) Choi et al.\ 2004; (7) Clarke et al.\ 2004; (8) Clarke et al.\ 2005; (9) David et al.\ 2001; (10) Durret et al.\ 2005; (11) Ettori et al.\ 2002; (12) Fabian et al.\ 2000; (13) Fabian et al.\ 2003a; (14) Fabian et al.\ 2005; (15) Fabian et al.\ 2006; (16) Finoguenov \& Jones 2001; (17) Forman et al.\ 2005; (18) Fujita et al.\ 2002; (19) Fujita et al.\ 2004; (20) Heinz et al.\ 2002; (21) Hicks et al.\ 2002; (22) Iwasawa et al.\ 2001; (23) Jeltema et al.\ 2005; (24) Johnstone et al.\ 2002; (25) Johnstone et al.\ 2005; (26) Kempner et al.\ 2002; (27) Kino \& Kawakatu 2005; (28) Kraft et al. 2006; (29) Leahy \& Grizani 2001; (30) Mazzotta et al.\ 2002; (31) Mazzotta et al.\ 2003; (32) Mazzotta et al.\ 2004; (33) McNamara et al.\ 2000; (34) McNamara et al.\ 2001; (35) McNamara et al.\ 2004; (36) McNamara et al.\ 2005; (37) McNamara et al.\ 2006; (38) Nulsen et al.\ 2002; (39) Nulsen et al.\ 2005a; (40) Nulsen et al.\ 2005b; (41) Reynolds et al.\ 2005; (42) Sanders \& Fabian 2002; (43) Sanders et al.\ 2005; (44) Sanderson et al.\ 2005; (45) Schindler et al.\ 2001; (46) Schmidt et al.\ 2001; (47) Smith et al.\ 2002; (48) Sun et al.\ 2003; (49) Vrtilek et al.\ 2002; (50) Wise et al.\ 2004; (51) Young et al.\ 2002; (52) Zakamska \& Narayan 2003.}
\tablenotetext{a}{Central stellar velocity dispersions were taken from the HyperLeda database; when more than one measurement was available, a weighted average was used. For the purposes of 
the buoyancy-age calculation, when no velocity dispersion was available, the average value for our sample ($\left< \sigma \right> = 295$ km~s$^{-1}$) was adopted.}
\tablenotetext{b}{Total magnitudes from the 2MASS catalog (\emph{K-}band) or HyperLeda catalog (\emph{R-}band), corrected for Galactic extinction, $K$-correction, and evolution (see text for details).}
\tablenotetext{c}{Bulge mass calculated from the \emph{R-}band absolute magnitude. Errors reflect uncertainties in $M_R$ only.}
\tablenotetext{d}{The \textit{Chandra} image of A1068 does not show evidence of cavities. A1068 is included because of the large starburst in the central galaxy.}
\end{deluxetable*}

\section{The Sample}
Our sample was drawn primarily from the \citet{birz04} sample of cooling flows whose central galaxies show evidence of AGN activity as revealed by cavities in X-ray images. We have supplemented this sample with a number of recently discovered cavity systems and one non-cavity system (A1068) with high-quality star formation data. In total, our sample comprises 31 CDGs, 1 group dominant galaxy (HCG 62), and 1 giant elliptical (M84). Table \ref{T:sample} lists the properties of the sample and references for publications that discuss the cavities or X-ray data. All the systems in our sample were observed with \textit{Chandra} and have data publicly available in the \textit{Chandra} Data Archive. The sample ranges in redshift from $z=0.0035$ to $z=0.545$ and varies in its composition from groups to rich clusters. We note that our sample is biased in favor of cavity systems; therefore, conclusions drawn from this sample may not apply to cooling flows as a whole.

\section{Data Analysis}\label{S:Analysis}
The following section describes the reduction and analysis of data used in this paper. Briefly, the cavities were identified and their sizes and projected radial distances were measured from \textit{Chandra} X-ray data. The temperature and density of the ICM as a function of radius, used to find the pressures at the projected locations of the cavities, were also derived from \textit{Chandra} data. The internal pressure of the cavities was derived under the assumption that the cavities are approximately in pressure balance with the surrounding medium.  The cavity's pressure, size, and position were then used to find the mean cavity power. Black hole growth rates were inferred from the cavity power under the assumption that accretion onto the central black hole fuels the outburst. Lastly, high-quality star formation and cooling rates were taken from the literature. We also use lower-quality cooling rates derived from \textit{Chandra} data. Unless otherwise noted, errors and upper limits are 1$\sigma$ values.

\subsection{\emph{Chandra} X-ray analysis}
All systems were observed with the \textit{Chandra} ACIS detector and data were obtained from the \textit{Chandra} Data Archive. The data were reprocessed with CIAO 3.3 using CALDB 3.2.0 and were corrected for known time-dependent gain and charge transfer inefficiency problems.  Blank-sky background files, normalized to the count rate of the source image in the $10-12$ keV band, were used for background subtraction.\footnote{See http://asc.harvard.edu/contrib/maxim/acisbg/} 
\vspace{1mm}

\subsubsection{Temperature and Density Profiles}\label{S:X-ray_analysis}
Spectra were extracted in elliptical annuli centered on the X-ray centroid of the cluster with eccentricity and position angle set to the average values of the cluster isophotes.  Weighted response files were made using the CIAO tools MKWARF and MKACISRMF or MKRMF (MKACISRMF was used for all observations taken at the $-120$ C focal plane temperature; MKRMF was used for all other observations).

Gas temperatures and densities were found by deprojecting the spectra with a single-temperature plasma model (MEKAL) with a foreground absorption model (WABS) using the PROJCT mixing model in XSPEC 11.3.2, between energies of 0.5 keV and 7.0 keV. The redshift was fixed to the value given in Table \ref{T:sample}, and the foreground hydrogen column density was fixed to the Galactic value of \citet{dick90}, except in the case of 2A 0335+096 and A478, when a significantly different value was required by the fit. In these two cases, the column density in each annulus was allowed to vary.

\def\arraystretch{1.3}
\begin{deluxetable*}{lcccccl}
\tabletypesize{\scriptsize}
\tablewidth{0pt}
\tablecaption{Cavity and ICM Properties Derived from \textit{Chandra} Data. \label{T:properties}}
\tablehead{
	\colhead{} & \colhead{$pV_{\rm{tot}}$} & \colhead{$P_{\rm{cav,tot}}$\tablenotemark{a}}  & \colhead{$L_{\rm{X}}$\tablenotemark{b}} &
	\colhead{$\dot{M}_{\rm{cool}}$\tablenotemark{c}} & \colhead{$L_{\rm{cool}}$\tablenotemark{b}} & \colhead{$r_{\rm{cool}}$\tablenotemark{d}}\\
	\colhead{System\phn} & \colhead{($10^{58}$ erg)} & \colhead{($10^{42}$ erg s$^{-1}$)} &	\colhead{($10^{42}$ erg s$^{-1}$)} & 
	\colhead{($M_{\sun}$ yr$^{-1}$)} & \colhead{(10$^{42}$ erg s$^{-1}$)} & \colhead{(kpc)}  }
\startdata
	A85	    & $1.2_{-0.4}^{+1.2}$       & $37_{-11}^{+37}$	& $365 \pm 20$		& $18_{-9}^{+13}$	& $30_{-10}^{+20}$	& 142	\\
	A133	    & $24_{-1}^{+11}$		& $620_{-20}^{+260}$	& $106 \pm 2$		& $5 \pm 3$		& $3 \pm 2$ 		& \phn93	\\
	A262        & $0.13_{-0.03}^{+0.10}$ 	& $9.7_{-2.6}^{+7.5}$	& $11.1_{-0.3}^{+0.4}$ 	& $<0.7$	  	& $<0.3$ 		& \phn57	\\
	Perseus     & $19_{-5}^{+20}$    	& $150_{-30}^{+100}$	& $554 \pm 2$ 		& $20_{-8}^{+9}$  	& $21_{-7}^{+8}$	& \phn90*	\\
   	2A 0335+096 & $1.1_{-0.3}^{+1.0}$    	& $24_{-6}^{+23}$ 	& $338 \pm 2$ 		& $29_{-5}^{+7}$        & $13 \pm 4$		& 135	\\
	A478        & $1.5_{-0.4}^{+1.1}$   	& $100_{-20}^{+80}$	& $1440 \pm 10$ 	& $40_{-20}^{+40}$	& $40_{-20}^{+50}$	& 150	\\ 
	MS 0735.6+7421& $1600_{-600}^{+1700}$	& $6900_{-2600}^{+7600}$& $450 \pm 10$		& $20_{-10}^{+20}$*	& $12_{-8}^{+13}$			& 141	\\
	PKS 0745-191& $69_{-10}^{+56}$   	& $1700_{-300}^{+1400}$ & $2300 \pm 30$		& $170 \pm 90$	  	& $230 \pm 120$		& 176	\\
	4C 55.16	    & $12_{-4}^{+12}$		& $420_{-160}^{+440}$	& $640 \pm 20$		& $70 \pm 30$		& $70 \pm 20$		& 162	\\
	Hydra A     & $64_{-11}^{+48}$    	& $430_{-50}^{+200}$    & $282 \pm 2 $	& $16 \pm 5$		& $13 \pm 4$		& 109 	\\
	RBS 797	    & $38_{-15}^{+50}$		& $1200_{-500}^{+1700}$	& $3100_{-130}^{+100}$	& $200_{-180}^{+490}$*	& $250_{-220}^{+400}$	& 185	\\
	Zw 2701	    & $350_{-200}^{+530}$	& $6000_{-3500}^{+8900}$& $430_{-30}^{+20}$	& $<8$*			& $<6$			& 135	\\
	Zw 3146	    & $380_{-110}^{+460}$       & $5800_{-1500}^{+6800}$& $3010_{-90}^{+70}$	& $590_{-170}^{+190}$   & $680_{-150}^{+170} $		& 186	\\
	A1068	    & \nodata  			& 20\tablenotemark{e}   & \nodata 		& $<48$	  		& \nodata		& 152	\\
	M84         & $0.003_{-0.002}^{+0.005}$ & $1.0_{-0.6}^{+1.5}$	& $0.07 \pm 0.01$ 	& $0.038 \pm 0.002$* 	& $0.012_{-0.001}^{+0.003}$ & \phn10	\\
	M87         & $0.020_{-0.003}^{+0.014}$ & $6.0_{-0.9}^{+4.2}$	& $8.30_{-0.04}^{+0.03}$& $1.2_{-0.3}^{+0.1}$   & $1.1_{-0.2}^{+0.1}$	& \phn26* 	\\
	Centaurus   & $0.060_{-0.015}^{+0.051}$ & $7.4_{-1.8}^{+5.8}$	& $28.1 \pm 0.3$	& $2.7_{-0.1}^{+0.2}$	& $4.3\pm 0.2$		& \phn54*	\\	
	HCG 62	    & $0.046_{-0.028}^{+0.073}$	& $3.9_{-2.3}^{+6.1}$	& $1.8 \pm 0.2$		& $<0.3$		& $<0.1$		& \phn33	\\
	A1795       & $4.7_{-1.6}^{+6.6}$    	& $160_{-50}^{+230}$	& $625_{-11}^{+6}$ 	& $8_{-7}^{+13}$	& $10_{-9}^{+14}$	& 135	\\
	A1835	    & $47_{-16}^{+50}$		& $1800_{-600}^{+1900}$ & $3160_{-90}^{+60}$	& \nodata		  	& \nodata			& 156	\\
	PKS 1404-267& $0.12_{-0.05}^{+0.15}$	& $20_{-9}^{+26}$	& $27 \pm 1$		& $5\pm 2$	 	& $3 \pm 1$		& \phn83	\\
	MACS J1423.8+2404  & $29_{-19}^{+52}$		& $1400_{-900}^{+2500}$	& $2290 \pm 30$		& $140_{-90}^{+110}$		& $90_{-60}^{+70}$		& 187	\\
	A2029       & $4.8_{-0.1}^{+2.7}$    	& $87_{-4}^{+49}$      & $1160 \pm 10$		& $<1.9$		& $<3$			& 140	\\
	A2052       & $1.7_{-0.7}^{+2.3}$    	& $150_{-70}^{+200}$	& $97\pm 1$ 		& $7.0_{-0.4}^{+0.9}$	& $3.4_{-0.2}^{+0.5}$	& \phn87	\\
	MKW 3S      & $38_{-4}^{+39}$    	& $410_{-44}^{+420}$    & $104 \pm 2$ 		& $5_{-2}^{+3}$		& $5_{-2}^{+3}$		& 120	\\
	A2199       & $7.5_{-1.5}^{+6.6}$    	& $270_{-60}^{+250}$ 	& $142 \pm 1$ 		& $<3$   		& $<3$			& \phn91	\\
	Hercules A  & $31_{-9}^{+40}$		& $310_{-90}^{+400}$	& $210_{-20}^{+10}$	& $<58$* 		& $<46$			& 104	\\
	3C 388	    & $5.2_{-2.1}^{+7.5}$	& $200_{-80}^{+280}$	& $27_{-3}^{+2}$	& $<3$*			& $<2$			& \phn55	\\
	3C 401	    & $11_{-7}^{+20}$		& $650_{-420}^{+1200}$	& $37_{-7}^{+2}$	& $12_{-6}^{+5}$*	& $7 \pm 3$		& \phn62	\\
	Cygnus A    & $84_{-14}^{+70}$		& $1300_{-200}^{+1100}$	& $420 \pm 4$		& $31_{-6}^{+7}$	& $50 \pm 10$		& \phn91	\\
	Sersic 159/03& $25_{-8}^{+26}$		& $780_{-260}^{+820}$	& $220 \pm 6$		& $15 \pm 9$		& $9 \pm 5$		& 136	\\
	A2597       & $3.6_{-1.5}^{+4.6}$    	& $67_{-29}^{+87}$      & $470_{-17}^{+8}$ 	& $30_{-20}^{+30}$  	& $30_{-20}^{+30}$	& 128	\\
	A4059       & $3.0_{-0.9}^{+2.5}$    	& $96_{-35}^{+89}$	& $93\pm 1$ 		& $3_{-2}^{+2}$  	& $2 \pm 1$		& \phn85	
\enddata
\tablenotetext{a}{Cavity power calculated assuming $4pV$ of energy per cavity and the buoyancy timescale.}
\tablenotetext{b}{Bolometric luminosity between 0.001 and 100 keV inside $r_{\rm{cool}}$.}
\tablenotetext{c}{Net cooling rate to low temperatures. Values marked with an asterisk where derived from observations with a low number of counts inside the cooling radius ($\lesssim 15000$) and are therefore less reliable.}
\tablenotetext{d}{Radius of the cooling region, inside which the cooling time is less than $7.7 \times 10^9$ yr (except for values marked with an asterisk, which correspond to the radius at the chip's edge).}
\tablenotetext{e}{For A1068, the cavity power was calculated from the $\nu=1400$ MHz radio power as $P_{\rm{cav,tot}} \sim 1500 \times \nu_{\rm{MHz}} P_{\nu}$.}
\end{deluxetable*}

\subsubsection{Cavity Power}\label{S:Cav_power}
Cavities seen in the X-ray emission of clusters allow a direct measurement of the non-radiative energy output via jets from the AGN. This measurement is independent of the radio properties and is the most reliable available, since its derivation rests on only a few, well-understood quantities. The radio-emitting jets are understood to be displacing the ICM at the location of the cavities, doing work against the surrounding plasma, as well as supplying thermal energy to the radio plasma that fills the lobes. The total energy required to create a cavity is equal to its enthalpy, given by
\begin{equation}
E_{\rm{cav}} = \frac{\gamma}{(\gamma - 1)} pV,
\end{equation}
where $p$ is the pressure of the gas surrounding the cavity, $V$ is the cavity's volume, and $\gamma$ is the ratio of specific heats of the gas inside the cavity.  For a relativistic gas, $\gamma=4/3$, and the enthalpy is $4pV$. We assume this value of $\gamma$ for all subsequent calculations involving $E_{\rm{cav}}$. 

The cavity's size and position were measured following the procedures used in \citet{birz04}.  The cavity was assumed to be in pressure equilibrium with its surroundings, and hence its pressure was taken to be the azimuthally averaged value at the projected radius of its center. The cavity's age was estimated in three ways: by assuming the cavity to be a buoyant bubble that rises at its terminal velocity, by assuming that the bubble moves outward at the local sound speed, and by assuming the cavity's age is governed by the time required for material to refill the volume of the cavity as it moves outward \citep[for a detailed description of our analysis, see][]{birz04}. These ages typically differ by factors of 2-4. For simplicity, we use the buoyancy age as the estimate of cavity's age. This estimate is probably an upper limit on the age of the cavity (neglecting projection effects), since the cavity is expected to move outward supersonically during the early, momentum-dominated phase of the jet. 

The mean jet power required to create a cavity or cavity pair is then 
\begin{equation}
P_{\rm{cav}} = \frac{E_{\rm{cav}}}{t},
\end{equation}
where $t$ is the average time between outbursts.  This time is known only for a few objects, such as Perseus, for which the interval between outbursts may be estimated from the presence of multiple generations of cavities and ripples \citep{fabi06}.  In objects with only a single set of cavities, which make up most of our sample, the cavity's buoyancy age is used for $t$. 

As noted above, the buoyancy age is likely to be an overestimate of the true age, and time evolution in the output of the AGN's jets can lead to underestimates of the amount of total energy traced by the cavities. Furthermore, the discovery of shocks in a number of deep X-ray images of clusters (e.g., Cygnus A, Wilson et al.\ 2003; NGC 4636, Jones et al.\ 2002; MS0735+7421, McNamara et al.\ 2005; Hercules A, Nulsen et al.\ 2005a), which typically represent a comparable amount of energy as that contained in the cavities, may mean that cavities trace a fraction ($\sim 50$\%) of the energy of a typical outburst. Lastly, our cavity powers do not include the radiative luminosity of the AGN. Therefore, our estimates of $P_{\rm{cav}}$ represent a lower limit to the total power of the AGN. 

Table \ref{T:properties} lists the total cavity energies and the associated powers for the systems in our sample (see Table \ref{T:cavities} in the Appendix for the properties of each cavity). For A1068, in which no cavities are apparent in the X-ray image, the outburst power was estimated using the $\nu=1400$ MHz radio flux from the NVSS survey ($S_{1400}=23.1\pm 1.1$ mJy), as $P_{\rm{cav,tot}} \sim 1500 \times \nu_{\rm{MHz}} P_{\nu},$ the average relation found from the sample of \citet{birz04} for radio-filled cavities. 

\subsubsection{X-ray and Cooling Luminosities}\label{S:Cooling_analysis}
As in \citet{birz04}, we wish to compare the cavity powers to the heating rates required to balance losses from the ICM due to X-ray emission. These losses may be estimated as the difference between the total X-ray luminosity and the luminosity of gas cooling to low temperatures (i.e. out of the X-ray band). In this analysis, we define the cooling radius as the radius (or semi-major axis if elliptical annuli were used) within which the gas has a cooling time less than $7.7\times 10^{9}$ yr (the time since $z=1,$ representative of the time that the cluster has been relaxed and a cooling flow could become established). For those systems in which the cooling radius lies beyond the chip's edge, we use the radius at the chip's edge as the cooling radius. 

The deprojection described in Section \ref{S:X-ray_analysis} was performed again, and the bolometric flux of the MEKAL component inside the cooling radius was used to calculate the X-ray luminosity, $L_{\rm{X}}$. The same model, with the addition of a cooling-flow component (MKCFLOW), was used to obtain an estimate of the net cooling rate and the associated cooling luminosity ($L_{\rm{cool}}$) of gas cooling to low temperatures (found by fixing the MKCFLOW low temperature to 0.1 keV). In the case of A1835, the spectra were of insufficient quality to obtain a reliable cooling rate \citep[see][]{mcna06}. Table \ref{T:properties} lists the luminosities and cooling rates derived from \textit{Chandra} data. We use these rates, and those from XMM-\textit{Newton} and FUSE (described in Section \ref{S:Net_cooling}), as estimates of the net cooling rate of the ICM, which we compare to the star formation rate of the central galaxy in Section \ref{S:SF_Cooling}. 

\subsection{Black Hole Growth Rates}\label{S:BH}
The energies and ages described in Section \ref{S:Cav_power} may be used to infer the minimum growth rate of the black hole assuming the cavities were created by AGN jets fueled by accretion onto the central black hole. Although the luminous energy radiated by the AGN is not included in the cavity energies and must also be fueled by accretion, these systems are radiatively inefficient \citep[e.g.,][]{birz04}, and the contribution of radiation to the current total power is negligible. The outbursts might pass through a radiatively efficient phase during their initial stages, but this phase could not have been long-lived, since cluster AGN do not now show the quasar-like activity, and should not therefore affect our results significantly. We stress that our black hole growth rates are lower limits; any energy in excess of the jet energy would result in underestimates of the average accretion rates, but we expect this effect to be small. 

The jets are produced through the partial conversion (with efficiency $\epsilon$) of the gravitational binding energy of the accreting material into outburst energy.  The energy required to create the cavities requires an accretion mass, $M_{\rm{acc}},$ of 
\begin{equation}\label{E:m_acc}
M_{\rm{acc}} = \frac{E_{\rm{cav}}}{\epsilon c^2}.
\end{equation}
The value of $\epsilon$ depends on poorly understood details of the jet production process and, probably, on black hole spin. Under the usual assumption, that the maximum energy that can be extracted is determined by the binding energy of the last stable orbit, the upper limit on the efficiency ranges from $\epsilon \lesssim 0.06$ for a nonrotating black hole to $\epsilon \lesssim 0.4$ for an extreme Kerr black hole \citep{king02}. We assume when calculating the energy of the outburst that each cavity represents $4pV$ of energy (i.e.\ that they are filled with a relativistic plasma). 

Since some of the accreting material's mass goes to power the jets, the black hole's mass grows by
\begin{equation}\label{E:m_bh}
\Delta M_{\rm{BH}} = (1-\epsilon)M_{\rm{acc}}.
\end{equation}
Therefore, increased efficiency results in smaller black hole growth for a given outburst energy. The time-averaged accretion and black hole growth rates were found by dividing Equations \ref{E:m_acc} and \ref{E:m_bh} by the characteristic time scale discussed in Section \ref{S:Cav_power}. Table \ref{T:BH_masses} lists the inferred mass by which the black hole grew and the average rate of growth during the outburst. The implied black hole growth rates vary across our sample by approximately four orders of magnitude, from $1.6 \times 10^{-4}$ $M_\sun$ yr$^{-1}$ (M87) to 1.1 $M_\sun$ yr$^{-1}$ (MS 0735.6+7421), with an average value of $\sim 0.1$ $M_\sun$ yr$^{-1}$ and a median value of 0.035 $M_\sun$ yr$^{-1}$.

\begin{deluxetable*}{lccccccc}
\tabletypesize{\scriptsize}
\tablewidth{0pt}
\tablecaption{Black Hole Masses and Growth Rates. \label{T:BH_masses}}
\tablehead{
	\colhead{} & \colhead{$M_{\rm{BH,meas}}$\tablenotemark{a}} & \colhead{$M_{\rm{BH,\sigma}}$} & \colhead{$M_{{\rm BH,}L_{K}}$\tablenotemark{b}} & \colhead{$\Delta M_{\rm{BH}}$\tablenotemark{c}} &
	\colhead{$\dot{M}_{\rm{BH}}$\tablenotemark{c}} & \colhead{Bondi ratio\tablenotemark{d}} & \colhead{Eddington ratio\tablenotemark{d}} \\
	\colhead{System\phn} & \colhead{($10^9$ $M_{\sun}$)} & \colhead{($10^9$ $M_{\sun}$)} & \colhead{($10^9$ $M_{\sun}$)} & 
	\colhead{($M_{\sun}$)} & \colhead{($M_{\sun}$ yr$^{-1}$)} & \colhead{$(\dot{M}_{\rm{acc}} / \dot{M}_{\rm{Bondi}})$} & \colhead{$(\dot{M}_{\rm{acc}} / \dot{M}_{\rm{Edd}})$} }
\startdata
	A85	    & \nodata & $1.1_{-0.4}^{+0.6}$	    & $1.0_{-0.4}^{+0.6}$ & 2.5$_{-0.7}^{+2.5}\times 10^{5}$ & $5.9_{-1.7}^{+5.9} \times 10^{-3}$	& 12$_{-9}^{+52}$	& $2.6_{-1.4}^{+5.2} \times 10^{-4}$	\\
	A133        & \nodata & \nodata 	    & $0.7_{-0.3}^{+0.4}$ & 5.0$_{-0.2}^{+2.1}\times 10^{6}$ & $9.8_{-0.4}^{+4.2} \times 10^{-2}$	& 1000$_{-670}^{+3300}$	& $7.3_{-2.9}^{+9.2} \times 10^{-3}$\\
	A262        & \nodata & $0.4_{-0.1}^{+0.2}$         & $0.3\pm0.1$ & 2.5$_{-0.7}^{+2.0}\times 10^{4}$ & $1.5_{-0.4}^{+1.2} \times 10^{-3}$ &  14$_{-9}^{+41}$ & $2.1_{-1.0}^{+3.2} \times 10^{-4}$\\
	Perseus     & \nodata & $0.3\pm0.1$	    & $0.6_{-0.2}^{+0.3}$ & 3.8$_{-1.0}^{+3.9}\times 10^{6}$ & $2.4_{-0.5}^{+1.5} \times 10^{-2}$ & 1400$_{-900}^{+4200}$  & $3.9_{-1.7}^{+5.2} \times 10^{-3}$\\ 
   	2A 0335+096 & \nodata & \nodata  	    & $0.5_{-0.2}^{+0.3}$ & 2.2$_{-0.6}^{+2.0}\times 10^{5}$ & $3.8_{-1.1}^{+3.7} \times 10^{-3}$ & 36$_{-26}^{+137}$  & $3.6_{-1.9}^{+7.1} \times 10^{-4}$\\
	A478        & \nodata & \nodata  	    & $0.9_{-0.4}^{+0.6}$ & 3.0$_{-0.7}^{+2.3}\times 10^{5}$ & $1.6_{-0.4}^{+1.2} \times 10^{-2}$ & 41$_{-32}^{+206}$  & $9.0_{-4.8}^{+17} \times 10^{-4}$\\ 
	MS 0735.6+7421& \nodata & \nodata		    & $0.7_{-0.3}^{+0.5}$ & 3.2$_{-1.2}^{+3.5}\times 10^{8}$	& $1.1_{-0.4}^{+1.1}$	& 18000$_{-15000}^{+103000}$	& $7.9_{-5.0}^{+20} \times 10^{-2}$	\\
	PKS 0745-191& \nodata & \nodata  	    & $1.1_{-0.4}^{+0.7}$ & 1.4$_{-0.2}^{+1.1}\times 10^{7}$ & $2.7_{-0.4}^{+2.2} \times 10^{-1}$ & 630$_{-490}^{+3570}$	& $1.2_{-0.6}^{+2.5} \times 10^{-2}$	\\
	4C 55.16    & \nodata & \nodata		    & $0.5_{-0.2}^{+0.3}$ & 2.4$_{-0.9}^{+2.5}\times 10^{6}$	& $6.7_{-2.5}^{+7.0} \times 10^{-2}$	&1200$_{-900}^{+5600}$	&$6.5_{-4.0}^{+15} \times 10^{-3}$	\\
	Hydra A     & \nodata & $0.9_{-0.4}^{+0.7}$ & $0.4_{-0.1}^{+0.2}$ & 1.3$_{-0.2}^{+1.0}\times 10^{7}$ & $6.8_{-0.9}^{+3.1} \times 10^{-2}$ & 210$_{-170}^{+1300}$ & $3.7_{-1.8}^{+5.6} \times 10^{-3}$	\\
	RBS 797     & \nodata & \nodata	       	    & \nodata	  & 7.5$_{-3.0}^{+10}\times 10^{6}$	& $1.9_{-0.9}^{+2.7} \times 10^{-1}$	& \nodata & \nodata\\  
	Zw 2701	    & \nodata & \nodata		    & $0.6_{-0.2}^{+0.4}$ & 7.2$_{-4.1}^{+11}\times 10^{7}$	& $1.0_{-0.4}^{+1.4}$	& 62000$_{-54000}^{+453000}$	& $8.2_{-6.0}^{+25} \times 10^{-2}$	\\
	Zw 3146	    & \nodata & \nodata	      	    & $2.6_{-1.3}^{+2.4}$ & 7.7$_{-2.3}^{+9.2}\times 10^{7}$ & $0.9_{-0.2}^{+1.1}$ & 370$_{-300}^{+3060}$ & $1.7_{-1.0}^{+5.4} \times 10^{-2}$	\\
	A1068	    & \nodata & \nodata  	    & $1.0_{-0.4}^{+0.6}$ & \nodata	       & $3.1 \times 10^{-3}$ & 6.4$_{-4.3}^{+18}$ & $1.6_{-0.6}^{+1.0} \times 10^{-4}$	\\
	M84         & $0.36$  & $0.7\pm0.2$	    & $0.12\pm0.03$ & 6.1$_{-3.3}^{+8.4}\times 10^{2}$ & $1.6_{-0.9}^{+2.3} \times 10^{-4}$ & 0.44$_{-0.33}^{+1.5}$	& $2.1_{-1.4}^{+4.5} \times 10^{-5}$	\\
	M87         & $3.3\pm0.7$& $1.2_{-0.3}^{+0.5}$  & $0.3\pm0.1$	& 4.1$_{-0.6}^{+2.9}\times 10^{3}$ & $1.0_{-0.2}^{+0.7} \times 10^{-3}$ & 0.04$_{-0.02}^{+0.08}$ & $1.5_{-0.4}^{+1.7} \times 10^{-5}$\\
	Centaurus   & \nodata & $0.4\pm0.1$	    & $0.3\pm0.1$ & 1.2$_{-0.3}^{+1.0}\times 10^{4}$ & $1.2_{-0.3}^{+0.9} \times 10^{-3}$ & 2.4$_{-1.5}^{+5.9}$ & $1.6_{-0.7}^{+2.2} \times 10^{-4}$	\\	
	HCG 62      & \nodata & \nodata	     	    & $0.2\pm0.1$ & 9.3$_{-5.5}^{+15}\times 10^{3}$ & $6.1_{-3.6}^{+9.7} \times 10^{-4}$ & 12$_{-10}^{+54}$	& $1.4_{-1.0}^{+3.7} \times 10^{-4}$	\\
	A1795       & \nodata & $0.6_{-0.2}^{+0.3}$	    & $0.8_{-0.3}^{+0.5}$ & 9.4$_{-3.1}^{+13}\times 10^{5}$ & $2.6_{-0.9}^{+3.6} \times 10^{-2}$ & 390$_{-300}^{+2660}$ & $2.0_{-1.1}^{+5.2} \times 10^{-3}$	\\ 
	A1835	    & \nodata & \nodata	      	    & $1.9_{-0.9}^{+1.6}$ & 9.5$_{-3.3}^{+10}\times 10^{6}$ & $2.8_{-1.0}^{+3.0} \times 10^{-1}$ & 520$_{-430}^{+3570}$ & $7.3_{-4.7}^{+20} \times 10^{-3}$	\\
	PKS 1404-267& \nodata & $0.4\pm0.1$	    & $0.2\pm0.1$ & 2.3$_{-1.0}^{+2.9}\times 10^{4}$ 	& $3.2_{-1.5}^{+4.1} \times 10^{-3}$	&72$_{-53}^{+266}$	&$4.3_{-2.6}^{+9.2} \times 10^{-4}$	\\
	MACS J1423.8+2404  & \nodata & \nodata		    & \nodata	  & 5.8$_{-3.8}^{+11}\times 10^{6}$	&$2.2_{-1.4}^{+4.0} \times 10^{-1}$	&\nodata	&\nodata	\\
	A2029       & \nodata & $1.5_{-0.5}^{+0.8}$ & $2.1_{-0.9}^{+1.6}$ & 9.6$_{-0.3}^{+5.4}\times 10^{5}$ & $1.4_{-0.1}^{+0.8} \times 10^{-2}$ & 7.0$_{-4.7}^{+26}$ & $4.4_{-1.7}^{+6.3} \times 10^{-4}$  \\
	A2052       & \nodata & $0.4_{-0.1}^{+0.2}$	    & $0.6_{-0.2}^{+0.3}$ & 3.6$_{-1.5}^{+4.5}\times 10^{5}$ & $2.3_{-1.1}^{+3.1} \times 10^{-2}$ & 510$_{-420}^{+2710}$ 	& $3.1_{-1.9}^{+7.6} \times 10^{-3}$	\\
	MKW 3S      & \nodata & \nodata 	    & $0.3\pm0.1$ & 7.7$_{-0.8}^{+7.9}\times 10^{6}$ & $6.5_{-0.7}^{+6.7} \times 10^{-2}$ & 12000$_{-9000}^{+98000}$ & $1.1_{-0.4}^{+2.2} \times 10^{-2}$	\\
	A2199       & \nodata & $0.7_{-0.2}^{+0.3}$ & $0.7_{-0.2}^{+0.4}$ & 1.5$_{-0.3}^{+1.3}\times 10^{6}$ & $4.3_{-1.0}^{+3.9} \times 10^{-2}$ & 260$_{-170}^{+860}$ & $3.1_{-1.3}^{+4.9} \times 10^{-3}$	\\
	Hercules A  & \nodata & \nodata 	    & $0.7_{-0.3}^{+0.5}$ & 6.3$_{-1.7}^{+8.1}\times 10^{6}$ & $5.0_{-1.4}^{+6.4} \times 10^{-2}$ & $2100_{-1600}^{+13100}$& $3.3_{-1.9}^{+9.2} \times 10^{-3}$ \\
	3C 388	    & \nodata & $1.5_{-0.7}^{+1.2}$ & $0.6_{-0.2}^{+0.3}$ & 1.1$_{-0.4}^{+1.5}\times 10^{6}$	&$3.1_{-1.2}^{+4.4} \times 10^{-2}$	&960$_{-810}^{+7750}$	& $1.0_{-0.7}^{+3.4} \times 10^{-3}$	\\
	3C 401	    & \nodata & \nodata		    & \nodata	  & 2.2$_{-1.4}^{+3.9}\times 10^{6}$	&$1.0_{-0.7}^{+1.9} \times 10^{-1}$	&\nodata	&\nodata	\\
	Cygnus A    & $2.7\pm0.7$& \nodata & $1.0_{-0.4}^{+0.6}$ & 1.7$_{-0.3}^{+1.4}\times 10^{7}$	& $2.1_{-0.4}^{+1.8} \times 10^{-1}$	& $210_{-120}^{+630}$	& $3.6_{-1.2}^{+5.2} \times 10^{-3}$	\\
	Sersic 159/03& \nodata& \nodata		    & $0.6_{-0.2}^{+0.4}$ & 5.0$_{-1.6}^{+5.3}\times 10^{6}$	& $1.2_{-0.4}^{+1.3} \times 10^{-1}$	& 1400$_{-1100}^{+8100}$	& $1.0_{-0.6}^{+2.4} \times 10^{-2}$	\\
	A2597       & \nodata & $0.2\pm0.1$ 	    & $0.3\pm0.1$ & 7.2$_{-3.0}^{+9.3}\times 10^{5}$ & $1.1_{-0.5}^{+1.4} \times 10^{-2}$ & 640$_{-540}^{+4100}$ & $2.6_{-1.7}^{+7.1} \times 10^{-3}$	\\
	A4059       & \nodata & $0.7_{-0.4}^{+1.0}$ & $1.0_{-0.4}^{+0.6}$ & 6.1$_{-1.7}^{+5.2}\times 10^{5}$ & $1.5_{-0.6}^{+1.4} \times 10^{-2}$ & 450$_{-410}^{+5830}$ & \phn\phn$1.2_{-0.9}^{+4.6} \times 10^{-3}$	
\enddata
\tablenotetext{a}{Black hole mass measured using gas kinematics. For Cygnus A, the value of Tadhunter et al.\ (2003) was adopted, adjusted to our adopted angular diameter distance of 224.2 Mpc. For M87, the average of the values of Harms et al.\ (1994) and Macchetto et al. (1997) was adopted, adjusted to a distance of 17.9~Mpc. For M84, the value of  Maciejewski \& Binney (2001) was adopted, adjusted to a distance of 15.2~Mpc.}
\tablenotetext{b}{Values have been adjusted by a factor of 0.35 (see text for details).}
\tablenotetext{c}{The change and rate of change in black hole mass were calculated assuming $\epsilon=0.1$.}
\tablenotetext{d}{The Bondi and Eddington rates were calculated with $M_{\rm{BH,meas}}$ when available. If no measured value exists, $M_{\rm{BH,\sigma}}$ was used, if available, and 
$M_{{\rm BH,}L_{K}}$ if not.}
\end{deluxetable*}

\subsection{Eddington and Bondi Accretion Rates}\label{S:Edd_Bondi_rates}
It is useful to compare the inferred accretion rates to two theoretical rates, the Eddington and Bondi accretion rates. The Eddington rate is indicative of the maximum likely (steady-state) rate of accretion under the assumption of spherical symmetry, and occurs when the gravitational force acting inward on the accreting material is balanced by the outward pressure of the radiation emitted by the accretion process. For a fully ionized plasma, the Eddington accretion rate is
\begin{equation}
\frac{\dot{M}_{\rm{Edd}}}{\mbox{$M_\sun$ yr$^{-1}$}} = 2.2 \epsilon^{-1} \left( \frac{M_{\rm{BH}}}{10^9 M_\sun} \right).
\end{equation}
This rate is a function only of the black hole mass (discussed in Section \ref{S:BH_masses}) and the assumed radiative efficiency, $\epsilon$. Table \ref{T:BH_masses} lists the Eddington ratios ($\dot{M}_{\rm{acc}}/\dot{M}_{\rm{Edd}}$) for our sample, calculated assuming $\epsilon=0.1$.

The Bondi rate \citep{bond52} sets the rate of accretion, assuming spherical symmetry, for a black hole with an accreting atmosphere of temperature ($T$) and density ($n_{\rm{e}}$) as
\begin{equation}
\frac{\dot{M}_{\rm{Bondi}}}{\mbox{$M_\sun$ yr$^{-1}$}} = 0.012 \times \left( \frac{n_{\rm{e}}}{\rm{cm}^{-3}} \right) \left( \frac{kT}{\rm{keV}} \right)^{-3/2} \left( \frac{M_{\rm{BH}}}{10^9 M_\sun} \right)^2.
\end{equation}
This accretion occurs within the Bondi radius, inside which the gas comes under the dominating influence of the black hole: 
\begin{equation}\label{E:Bondi_radius}
\frac{R_{\rm{Bondi}}}{\mbox{kpc}} = 0.031 \times \left( \frac{kT}{\rm{keV}} \right)^{-1} \left( \frac{M_{\rm{BH}}}{10^9 M_\sun} \right).
\end{equation}
The Bondi rate is therefore an estimate of accretion directly from the hot ICM onto the black hole. Table \ref{T:BH_masses} lists the Bondi ratios ($\dot{M}_{\rm{acc}}/\dot{M}_{\rm{Bondi}}$) for our sample, and Table \ref{T:Bondi} in the Appendix lists the properties used in the calculation of the Bondi rates. In calculating the Bondi rate, we use the modeled temperature and density from \textit{Chandra} spectra, extracted from a central region that contains $\sim 3000$ counts after the exclusion of any non-thermal point sources. However, the size of the central region is not sufficiently small to resolve the Bondi radius of any system in our sample; therefore, the true temperature and density of the ICM at the Bondi radius could be lower and higher, respectively, than we have measured, resulting in an underestimate of the Bondi rate. We discuss this effect further in Section \ref{S:Accretion}.

\subsection{Black Hole Masses}\label{S:BH_masses}
Calculation of both the Eddington and Bondi rates requires estimates of the black hole mass. Of the systems in our sample, only three (Cygnus A, M84, and M87) have direct mass measurements (see Table \ref{T:BH_masses}). For the remaining systems, we use the bulge properties of the host galaxy as proxies for the black hole mass. As discussed earlier, the black hole's mass scales with the large-scale properties of the host galaxy such as bulge velocity dispersion and luminosity. The most well-studied relation between the black hole mass and the properties of the host galaxy is the $M_{\rm{BH}} - \sigma$ relation, which relates $M_{\rm{BH}}$ to the stellar velocity dispersion ($\sigma$) of the galaxy's bulge as 
\begin{equation}
\log \left( \frac{M_{\rm{BH,}\sigma}}{M_{\sun}} \right) = \alpha + \beta \log \left( \frac{\sigma}{\sigma_{0}} \right),
\end{equation}
where $\alpha,$ $\beta,$ and $\sigma_{0}$ are constants. The values of these constants vary somewhat from study to study \citep[for a discussion, see][]{trem02}. For the purposes of our calculations, we adopt the values of \citet{trem02}, namely $\alpha=8.13 \pm 0.06,$ $\beta=4.02 \pm 0.32,$ and $\sigma_{0}=200$~km s$^{-1}$. 

In deriving this relation, \citet{trem02} use as $\sigma$ the mean stellar velocity dispersion within a slit aperture of length $2 r_{\rm{e}}$ and width $1\arcsec - 2\arcsec$ (denoted by $\sigma_{1}$). Unfortunately, most of our sample lacks dispersions measured in this aperture. Instead, central velocity dispersions (generally measured within an aperture of $r \sim 2 \arcsec$) are more common.  Central dispersions (denoted by $\sigma_{\rm{c}}$) were taken from the HyperLeda Database.\footnote{Available at http://leda.univ-lyon.fr/} Measurements of $\sigma_{\rm{c}}$ exist for 15 of the 33 galaxies in our sample (listed in Table~\ref{T:sample}). When more than one measurement exists, we use the weighted average of all available measurements. We have estimated the magnitude of the error resulting from our use of $\sigma_{\rm{c}}$ instead of $\sigma_{1}$ using the relations given in \citet{jorg95} and \citet{trem02} and find for the 8 systems in our sample with measurements of both $r_{\rm{e}}$ and $\sigma_{\rm{c}}$ that $M_{\rm{BH,\sigma}}$ increases on average by 10\% after the correction, much less than the typical formal uncertainties in $M_{\rm{BH,\sigma}}$. Since we lack measurements of $r_{\rm{e}}$ for some systems and the correction is small, we ignore the aperture correction and use simply $\sigma_{\rm{c}}$ in our calculation of $M_{\rm{BH,\sigma}}$.

For the 18 systems without a measurement of velocity dispersion, we calculate the black hole mass from the total $K$-band luminosity of the bulge ($L_{K}$) using the relation of \citet{marc03} for their group 1 black holes (those with secure mass determinations):
\begin{equation}
\log \left( \frac{M_{{\rm BH,}L_K}}{M_{\sun}} \right) = A + B  \left[ \log \left( \frac{L_{K}}{L_{\sun}} \right) - 10.9 \right] ,
\end{equation}
where $A=8.21 \pm 0.07$ and $B=1.13 \pm 0.12$. Apparent $K$-band magnitudes were taken from the Two Micron All Sky Survey (2MASS) catalog.\footnote{See http://www.ipac.caltech.edu/2mass.} The apparent magnitudes were corrected for Galactic extinction with the values of \citet{schl98} and corrected for redshift ($K$-corrected) and evolution using the corrections of \citet{pogg97}.  Lastly, the magnitudes were converted to absolute magnitudes using our assumed cosmology and the redshifts listed in Table \ref{T:sample}.  

We note that there is a systematic offset between the masses calculated by the two methods for the 15 systems that have measurements of both central velocity dispersion and total $K$-band magnitude. Masses calculated from the total $K$-band luminosity are on average $2.9\pm1.6$ times greater. We checked this result using total $R$-band magnitudes from the HyperLeda database (see Section \ref{S:bulge}) and the $M_{\rm{BH}} - M_{R}$ relation of \citet{mclu04} and find a similar systematic offset [$M_{{\rm BH,}M_R} = (3.3\pm2.4) \times M_{\rm{BH,\sigma}}$]. \citet{bett03} find a similar but smaller offset in a sample of radio galaxies and attribute it to systematically low values of $\sigma$. Since out values of $\sigma$ are typically weighted averages of several values from a number of different sources, it is unlikely that they would be systematically low across our entire sample. 

We do not understand the origin of the offset in our data, but note that the galaxies in our sample are mostly large cDs, with extended stellar envelopes that may bias their total magnitudes with respect to normal ellipticals \citep[e.g.,][]{scho86}; however, \citet{fuji04a} do not find evidence of such an offset in a similar sample of CDGs.  It is also possible that the $M_{\rm{BH}} - \sigma$ relation breaks down at high masses \citep[see e.g.,][]{shie06}; however, there is little evidence to support this hypothesis at this time. \citet{marc03} find evidence of a significant correlation between $M_{\rm{BH}}$ and the bulge effective radius, with the result that $M_{\rm{BH,\sigma}}$ may be too low for large bulges. For typical values of the effective radius for galaxies in our sample ($r_{\rm{e}} \sim 10$ kpc), the magnitude of this effect is sufficient to account for the offset we see. However, \citet{marc03} note that this correlation is weak, and further investigation is required to confirm its existence. For the purposes of calculating the Eddington and Bondi rates, we adjust the black hole masses inferred from the $K$-band luminosities by a factor of 0.35. The black hole masses inferred by both methods are listed in Table \ref{T:BH_masses}.

\def\arraystretch{1.1}
\begin{deluxetable*}{lccccccc}
\tabletypesize{\scriptsize}
\tablewidth{0pt}
\tablecolumns{7}
\tablecaption{Star Formation and Cooling Rates. \label{T:SFR}}
\tablehead{
	\colhead{} & \multicolumn{3}{c}{Star Formation Rates} & \colhead{} & \multicolumn{3}{c}{Cooling Rates} \\
	\cline{2-4}  \cline{6-8} \\
	\colhead{} & \colhead{Continuous (ref)\tablenotemark{a}} & \colhead{Burst (ref)\tablenotemark{b}} & \colhead{Aperture} &
	\colhead{} & \colhead{XMM RGS (ref)\tablenotemark{c}} & \colhead{FUSE (ref)\tablenotemark{d}} & \colhead{Aperture} \\
	\colhead{System} & \colhead{($M_{\sun}$ yr$^{-1}$)} & \colhead{($M_{\sun}$ yr$^{-1}$)} & \colhead{(kpc)} & 
	\colhead{} & \colhead{($M_{\sun}$ yr$^{-1}$)} & \colhead{($M_{\sun}$ yr$^{-1}$)} & \colhead{(kpc)} 
	}
\startdata
	A262        & $<0.015$ (8)     &	\nodata			& $r=0.9$		& & $<2$ (15)	    & \nodata	& $r=5$	\\
	Perseus     & 2.3$\pm$0.2 (8)        &	\nodata			& $r=1.0$		& & \nodata	    & $32 \pm 6$ (3)& $11 \times 11$   \\	 
	            & 15.5$\pm$5.2 (16)      &	\nodata			& $r=18.8$		& &\nodata	    &\nodata	&\nodata  \\
	            & $\sim 37$ (18)	     &  \nodata			& $r=59$		& &\nodata	    &\nodata	&\nodata  \\
   	2A 0335+096 & 4.2 (17)               &	\nodata			& $r=16.0$		& & $20 \pm 10$ (15) &\nodata	& $r=22$	\\
 	A478        & 10.0 (4)               &	\nodata			& $3.1 \times 44.6$	& &\nodata	    &	\nodata&\nodata	  \\ 
	PKS 0745-191& 16.9$\pm$5.6 (16)      &	\nodata			& $r=18.8$		& &\nodata	    &\nodata	&\nodata  \\
	Hydra A     & $\lesssim 0.5$ (7)     &	$\sim 16$ (7)		& $r=4.3$ 		& & $35 \pm 20$ (15) &\nodata	& $r=22$	\\
	Zw 3146	    & 10.7 (5)        	     &	\nodata			& $5.7 \times (< 25.8)$ & &\nodata	    &\nodata	&\nodata  \\
		    & $<110$	(FIR)	     &  \nodata		  	& \nodata		& &\nodata	    &\nodata	&\nodata  \\
	A1068	    & 18.1 (5)		     &  \nodata			& $3.2 \times 15.6$	& &\nodata	    &\nodata	&\nodata  \\
		    & 28$\pm$12 (10)          &	46$\pm$21 (10)		& $r=10.0$		& &\nodata	    &\nodata	&\nodata  \\
	M84         & $<0.047$ (FIR)  &	\nodata			& \nodata		& & \nodata	    & $0.32$ (2)& $2.2 \times 2.2$   \\
	M87         & $<0.02$ (8)    &	\nodata			& $r=0.2$		& &$\lesssim 0.6$ (15) &\nodata	& $r=3.0$	\\
		    & $<0.081$ (FIR)  &	\nodata			& \nodata		& &\nodata	    &$\lesssim 0.44$ (2)& $2.6 \times 2.6$\\  
	A1795       & 0.95$\pm$0.10 (8)      &	\nodata			& $r=3.2$ 		& &\nodata	    & $26 \pm 7$ (3)& $36 \times 36$\\
	            & 1.1 (5)                &	\nodata			& $1.6 \times 7.6$      & &\nodata	    & $<15$ (14)	& $36 \times 36$	\\
	            & 2.1$\pm$0.9 (13)	     &  \nodata	& $19.4 \times 19.4$    & & $<30$ (15)  	    &\nodata	& $r=33$	\\
	            & 6.3 (9)		     &	23.2 (9)		& $13.3 \times 26.7$	& &\nodata	    &\nodata	&\nodata  \\
	A1835	    & 48.9 (5)       	     &	\nodata			& $5.1 \times 8.3$      & & $<200$ (15) 	    &\nodata	& $r=99$	\\
		    & 79.0 (5)       	     &	\nodata			& $5.1 \times 8.3$      & &\nodata	    &\nodata	&\nodata  \\
		    & 79.5 (5)	             &	\nodata			& $5.1 \times (< 23.7)$ & &\nodata	    &\nodata	&\nodata  \\
		    & 140$\pm$40 (11) 	     &	\nodata			& $r=30$		& &\nodata	    &\nodata	&\nodata  \\
	A2029       & $<0.15$ (8)            &	\nodata			& $r=3.9$		& &\nodata	    & $<27$ (3)	& $44 \times 44$	\\
	A2052       & 0.08$\pm$0.02 (8)      &	\nodata			& $r=1.8$		& & $<10$ (15)	    &\nodata	& $r=17$	\\
	            & 0.51 (5)               &	\nodata			& $0.9 \times 6.8$      & &\nodata	    &	\nodata &\nodata \\
	            & 0.31$\pm$0.10 (1)	     &  \nodata			& $r=2.1$	        & &\nodata	    &\nodata	&\nodata  \\
	MKW 3S      & $<0.03$ (8)            &	\nodata			& $r=2.3$		& & $<10$ (15)	    &\nodata	& $r=11$	\\
	A2199       & 0.10$\pm$0.03 (8)      &	\nodata			& $r=1.6$		& &\nodata	    &	\nodata&\nodata  \\
		    & 0.10 (5)               &	\nodata			& $0.8 \times 7.3$      & &\nodata	    &\nodata	&\nodata  \\
	A2597       & 2.3$\pm$1.3 (13)        &  \nodata		& $16.0 \times 16.0$    & & $\lesssim 50$ (12)	    & \nodata	& $r=190$ \\
		    & 6.4 (9)		     &	22.3 (9)		& $24.2 \times 24.2$	& &\nodata	    & 22 (14)	& $48 \times 48$	\\
	A4059      & \nodata           &	\nodata			& \nodata	& & $<10$ (15)	    &\nodata	& \phn\phn\phn$r=147$	    
\enddata
\tablenotetext{a}{Continuous star formation rate. References are in parentheses.}
\tablenotetext{b}{Star formation rate for a burst of star formation calculated as the mass of the burst divided by its age. References are in parentheses.}
\tablenotetext{c}{Cooling rates derived from XMM-\textit{Newton} RGS spectra. References are in parentheses.}
\tablenotetext{d}{Cooling rates derived from FUSE spectra. References are in parentheses.}
\tablerefs{ (1) Blanton et al.\ 2003; (2) Bregman et al.\ 2005; (3) Bregman et al.\ 2006; (4) Cardiel et al. 1998; (5) Crawford et al.\ 1999; (6) Lecavelier des Etangs et al.\ 2004; (7) McNamara 1995; (8) McNamara \& O'Connell 1989; (9) McNamara \& O'Connell 1993; (10) McNamara et al.\ 2004; (11) McNamara et al.\ 2006; (12) Morris \& Fabian 2005; (13) O'Dea et al.\ 2004; (14) Oegerle et al.\ 2001; (15) Peterson et al.\ 2003; (16) Romanishin 1987; (17) Romanishin \& Hintzen 1988; (18) Smith et al.\ 1992. }
\end{deluxetable*}

\subsection{Star Formation Rates}\label{S:SFR_analysis}
The determination of reliable star formation rates requires sensitive photometry over a broad wavelength range to identify and isolate the star-forming population. Secure star formation rates are available in the literature for a significant number of CDGs. We have collected these rates from the literature, adjusted to our assumed cosmology, and their sources in Table \ref{T:SFR}. Our sample includes rates derived from both spectroscopic and imaging studies. Readers wishing to skip the technical details should go directly to Section \ref{S:BH_bulge}.

Typically, in deriving star formation rates, one first finds the luminosity of the star-forming population. From broadband images, this luminosity may be found by modeling and subtracting a smooth background galaxy \citep[see e.g.,][]{mcna04}. Any extended excess emission is then assumed to be due to active star formation, and the resulting colors may then be compared to stellar population models to constrain its age and mass-to-light ratio (however, the age and mass-to-light ratio cannot be constrained unambiguously using colors alone). For spectra, a similar process is used whereby a spectrum of the background galaxy is subtracted (or included as a component in the models), and the remaining spectral features are then fit with stellar population models \citep[see e.g.,][]{craw99}. The models constrain the mass-to-light ratio and age of the star forming population, which may be used, together with its luminosity, to calculate the mean star formation rate. In both cases, the derived quantities are valid only in the aperture used. Consequently, there are three main sources of inhomogeneity in the star formation rates in our sample: differences and discrepancies in the model parameters (e.g., assumed ages), differences in the apertures within which the star formation is measured, and uncertainties due to dust extinction and reddening. 

There are two principle parameters that go into the stellar population synthesis models: the slope of the initial mass function (IMF) and the star formation history.  Changes in either of these parameters can result in typical deviations of factors of $\sim$5-10 in the derived star formation rates. For the systems in our sample that have star formation rates available, we list in Table~\ref{T:SFR} rates derived assuming continuous star formation for $\sim 10^9$ yr and, when available, for shorter duration bursts. There is little variation across our sample in IMF slope, since most studies assume a Salpeter IMF. This assumption appears to be valid in cooling flows \citep[e.g.,][]{mcna06}.

A significant difference between the studies we considered is the choice of aperture size. Observations made in spectroscopic slits have the weakness that the star forming region may not fall entirely within the slit, resulting in an underestimate of the total star formation rate. Table~\ref{T:SFR} gives the aperture used in each study. In our sample, aperture effects could lead to an underestimate of the total star formation rate by as much as a factor of $\sim 10$ if the star formation is uniformly distributed across the galaxy. However, imaging studies of CDGs \citep[e.g.,][]{mcna92,card98} show that star formation is centrally concentrated in most systems, reducing somewhat the likely magnitude of this effect. A comparison of objects in our sample with star formation rates derived in both ways shows that spectroscopic rates are typically lower than imaging-derived rates by factors of several. Therefore, although spectroscopic estimates should be treated as lower limits to the total star formation rates, they are unlikely to be more than an order of magnitude lower than the total rates.

In addition to observational and modeling inhomogeneities across our sample, a number of uncertainties exist in any derivation of the star formation rate. Principal among these are the effects of extinction and reddening due to dust. These effects are difficult to quantify without high resolution imaging which is not generally available. But comparison between the $U$-band rates, which are subject to strong extinction, and far IR rates, which are not, agree to within a factor of two \citep[e.g.,][]{mcna04,mcna06}. Of the objects in our sample, A2052 \citep{blan03}, A1068 \citep{mcna04}, 2A~0335+096 \citep{roma88}, A1795 and A2597 \citep{odea04}, and those systems studied by \citet{craw99} have published rates that have been corrected for the presence of dust. The intrinsic color excess for systems similar to those in our sample is typically $E(B-V) \sim 0.3$ \citep{craw99}. 

Lastly, errors in mass-to-light ratio and age, while leading to errors in accreted mass, generally result in robust star formation rates due to the compensating effect that older populations have higher mass-to-light ratios. Therefore, errors resulting from an overestimated age will be partly compensated by an overestimated population mass, reducing the error in the resulting star formation rate. 

A number of objects in our sample have no published optical star formation rates, or their rates were measured only in small apertures. For these objects, when possible, we have inferred the star formation rate from the far infrared (FIR) \textit{IRAS} 60 $\mu$m flux derived with the Infrared Processing and Analysis Center's SCANPI tool.\footnote{See http://irsa.ipac.caltech.edu/Missions/iras.html}  We used the following relation of \citet{kenn98} to convert the total FIR luminosity to a star formation rate:
\begin{equation}
\frac{SFR}{M_{\sun}~\rm{yr^{-1}}} \lesssim 4.5 \times  \left( \frac{\emph{L}_{\rm{FIR}}}{\rm{10^{44}~erg~s^{-1}}} \right),
\end{equation}
where $L_{\rm{FIR}} \sim 1.7 L_{60~\mu{\rm m}}$ \citep{rowa97}, and we have assumed that all the UV photons emitted by young stars are absorbed and re-radiated by dust in the FIR. Three objects in our sample have reliable 60 $\mu$m fluxes: Zw 3146, M87, and M84; for these objects we derived upper limits to the star formation rates (see Table \ref{T:SFR}).

\subsection{Bulge Masses}\label{S:bulge}
Lastly, to estimate the impact of star formation on the mass of the galaxy's bulge, we have estimated the mass of the bulge as
\begin{equation}
M_{\rm{bulge}} = L_{\rm{bulge}} \left( \frac{M}{L} \right)_{\rm{bulge}}.
\end{equation}
We use the total $R$-band luminosity of the galaxy for the bulge luminosity, $L_{\rm{bulge}}$. Total apparent magnitudes were taken when available from the catalog of \citet{prug98} and otherwise from the LEDA database (both databases are available through HyperLeda). In cases in which the $R$-band magnitudes were unavailable, we used the total $B$-band magnitudes, if available, and converted these to $R$-band magnitudes using $<(B-R)_{0}> = 1.44\pm0.17$ (the average corrected color of our sample). The apparent magnitudes were corrected and converted to absolute magnitudes in the same way as the $K$-band magnitudes (see Section \ref{S:Edd_Bondi_rates}). For the $R$-band mass-to-light ratio of the bulge, we adopt $(M/L)_{\rm{bulge}} = 6.3$ ($M/L)_{\sun},$ the average found by \citet{fish95}, after adjusting to our cosmology. The derived absolute magnitudes and bulge masses are listed in Table~\ref{T:sample}.

\subsection{Net X-ray and Far UV Cooling Rates}\label{S:Net_cooling}
Gas cooling out of the ICM at $T \sim 10^{7}$ K loses its energy primarily through thermal emission in the soft X-ray band. Therefore, its rate of cooling is best measured from X-ray spectra. Grating observations made with XMM-\textit{Newton} provide high spectral resolution and hence the best constraints on the cooling rates. \citet{pete03} have derived cooling rates from XMM-\textit{Newton} grating observations for nine of the objects in our sample, and we list the most constraining rate (i.e.\ the smallest rate in any of the temperature bands) from this study in Table~\ref{T:SFR}. With the exception of Hydra~A and 2A 0335+096, these rates are upper limits. We also list the rate derived for A2597 by \citet{morr05}.

At lower temperatures ($T \sim 10^{5}$ K), cooling gas should emit strongly in the far ultraviolet, mainly through the OVI doublet \citep[see][]{edga86}, where high-quality spectroscopic observations can be made. Such emission has been detected by FUSE for an additional six objects in our sample. The inferred FUSE cooling rates, calculated assuming the OVI emission is due to cooling gas \citep[see e.g.,][]{oege01,breg05,breg06}, are also listed in Table~\ref{T:SFR}.

\begin{figure}
\plotone{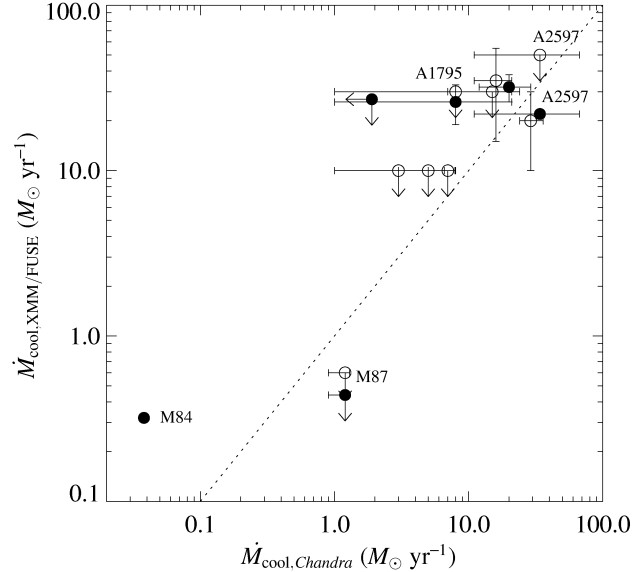}
\caption{Cooling rates derived from XMM-\textit{Newton} (\emph{empty symbols}) and FUSE (\emph{filled symbols}) spectra versus those derived from \textit{Chandra} spectra. M87, A1795, and A2597 have both XMM-\textit{Newton} and FUSE rates and hence appear twice. M84 is listed in \citet{breg05} as a probable FUSE detection. The dotted line denotes equality between the two rates.  \label{F:Cooling}}
\end{figure}

We have also derived cooling rates from lower-spectral-resolution \textit{Chandra} data (see Section \ref{S:Cooling_analysis}). For comparison, we plot the cooling rates from XMM-\textit{Newton} and FUSE against those from our \textit{Chandra} analysis in Figure \ref{F:Cooling}. Despite significant differences in aperture and in the details of the modeling, the agreement between the X--ray-derived rates is reasonably good, as is their agreement with the UV-derived FUSE rates. We note, however, that the \textit{Chandra} rates appear to be systematically lower than the other two rates, possibly due to spatial and spectral resolution effects or calibration and modeling differences.  

It should be emphasized that neither the \textit{Chandra} nor XMM-\textit{Newton} rates are based on fits to emission lines from gas cooling to low temperatures; rather, they are both based on fits to the continuum. Additionally, models fit to X-ray spectra do not generally require a cooling component to obtain an adequate fit. Therefore, X--ray-derived cooling rates should be interpreted as the \emph{maximum rates of cooling consistent with the spectra and not as unequivocal detections of cooling}. Until line emission that is uniquely due to cooling below 1 keV is identified, cooling through this temperature at any level cannot be confirmed \citep[see however][who find possible weak detections of several cooling lines in XMM-\textit{Newton} data of A2597]{morr05}. However, the reasonably close correspondence between FUSE and X--ray-derived rates indicates that cooling is occurring at or just below the detection limits.

\section{Results and Discussion}
\subsection{Black Holes and Bulges: Simultaneous Growth}\label{S:BH_bulge}
X-ray cavities provide a strong lower limit on the energy of the AGN outburst, independent of accretion disk radiation models and photon conversion efficiencies. Therefore they provide a robust means of estimating the minimum mass accreted onto the black hole. These properties allow us to investigate the relationship between the black hole's growth and the bulge's local (small-scale) growth in the same systems in a unique and detailed fashion that has not been possible before.

\begin{figure}
\plotone{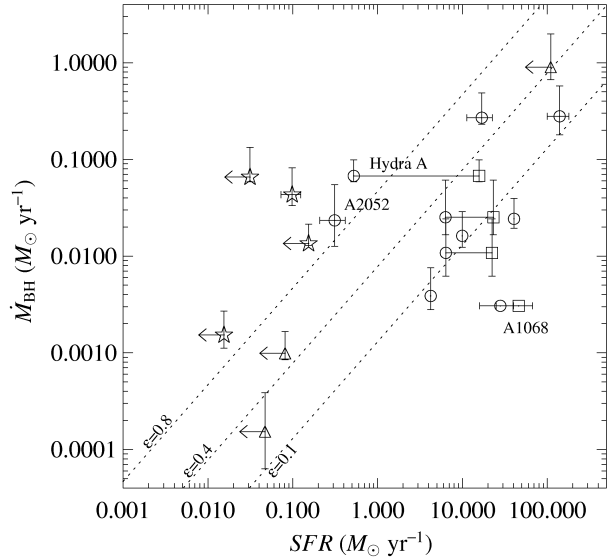}
\caption{Black hole growth rate versus star formation rate. The diagonal lines represent $\dot{M}_{\rm{BH}}=1.4 \times 10^{-3} SFR$ (see text for details) for different values of $\epsilon$. Circles denote continuous SFRs measured from broadband images, stars denote continuous SFRs measured from spectra taken in slits, and triangles denote continuous FIR SFRs. When more than one rate is available, we plot the largest rate. If an object has both broadband and spectral rates, we plot only the broadband rate. Squares denote rates for a burst of star formation and are joined to symbols denoting continuous rates for the same object by horizontal lines.\label{F:SFR_Mdot}}
\end{figure}

Figure \ref{F:SFR_Mdot} shows the black hole growth rate versus the bulge growth rate (traced by star formation) for the systems in our sample with reliable star formation rate estimates. We plot as dashed lines the time derivative of the present-day Magorrian relation, as found by \citet{hari04}: $\dot{M}_{\rm BH} = 1.4 \times 10^{-3} \dot{M}_{\rm bulge}$. In terms of our derived quantities, this relation becomes $(1-\epsilon) P_{\rm cav} /(\epsilon c^2) = 1.4 \times 10^{-3} SFR$. 

There is a trend, with large scatter, between the bulge and black hole growth rates, centered approximately on the Magorrian slope (assuming accretion efficiencies of $\epsilon \sim$ 0.1-0.4). 

As discussed in Section \ref{S:BH}, the upper limit on the efficiency with which the rest mass energy of the accreting material is converted to outburst energy varies between 0.06 and 0.4. Therefore, under the assumption that star formation traces all of the bulge's growth, consistency with general relativity and the slope of the Magorrian relation requires that all objects with estimates of the \emph{total} star formation rate fall below the $\epsilon=0.4$ line in Figure \ref{F:SFR_Mdot}. 

This requirement is clearly violated in a number of objects. For example, the black holes in both Hydra~A and A2052 are growing faster than strict adherence to the Magorrian relation would predict, whereas the bulge of A1068 is growing too fast (note, however, that no cavities were detected in A1068's atmosphere; therefore, there is large uncertainty in the rate of growth of A1068's black hole). While the discrepancy in A1068's rates may be explained with an extremely low  efficiency for the conversion of gravitational binding energy of the accreting material to outburst energy ($\epsilon \sim 0.005$), it is also possible that present-day growth is occurring in spurts, with periods of cooling and star formation (as in A1068) in which the bulge grows quickly with little commensurate black hole growth, while during periods of heating (as in Hydra A) the black hole grows more quickly than the bulge.

The trend in Figure \ref{F:SFR_Mdot} may be interpreted as an indication that, in a time-averaged sense, the growth of the bulges and black holes in our sample proceeds roughly along the Magorrian relation. When compared to the bulge masses calculated in Section \ref{S:SFR_analysis} and the black hole masses calculated in Section \ref{S:BH}, the black holes are growing at rates of $\sim 10^{-9} - 10^{-12}$ yr$^{-1}$ and the bulges at rates of $\sim 10^{-11} - 10^{-13}$ yr$^{-1}$. Present-day growth would not move most of the systems significantly off of the Magorrian relation, even if growth at such rates was constant for the age of the universe. 

However, for a number of systems, current growth could produce their present-day black holes in $\lesssim 10^{10}$ yr. The three most extreme cases (MS 0735.6+7421, Zw 2701, and Zw 3146) have growth rates that, if constant over just $\sim 10^9$ yr, would be sufficient to grow their black holes to their current masses. Periodic and powerful outbursts, without commensurate bulge growth (e.g., MS 0735.6+7421), could cause significant departures from the Magorrian relation.

These three systems represent $\sim 10$\% of our sample, implying a duty cycle in active systems of one such outburst every $\sim 10^8/0.1 = 10^9$ yr. Large outbursts might shut off cooling (and hence fueling) for long periods, making them a relatively rare event \citep[see][]{dona06}. If the most powerful outbursts are infrequent in the present-day universe, the Magorrian relation must have been established during earlier periods of extreme black hole and bulge growth, as has been postulated by a number of authors \citep[e.g.,][]{yu02,binn05,dima05,chur05} and supported by high-redshift quasar studies \citep[e.g.,][]{mclu04a}.  

Lastly, it is possible that we are missing some fraction of the bulge growth. The CDG is thought to grow through the addition of material by two main processes: cooling of gas out of the ICM (see Section \ref{S:X-ray_analysis}) and merging (cannibalism) of the CDG with other cluster members. The rate of growth from mergers is difficult to measure.  \citet{laue88}, through a study of multiple-nucleus CDGs, estimated a cannibalism rate of $L \approx 2L^{\star}$ per $5 \times 10^9$ yr.  This estimate implies that such growth is significant over the age of the cluster; however, the time scale for this growth is much longer than the cooling and star formation time scales considered here. Therefore, we have neglected mergers and used the star formation rates described in Section \ref{S:SFR_analysis} to set the instantaneous bulge growth rate.

\subsection{Accretion Mechanism}\label{S:Accretion}

\begin{figure}
\plotone{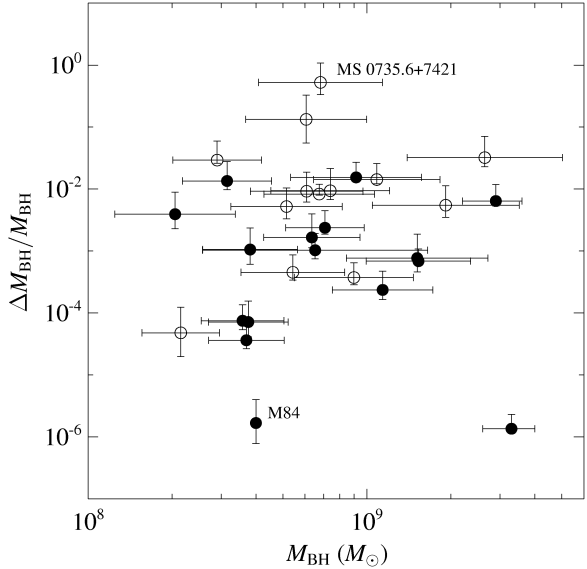}
\caption{The black hole's relative change in mass versus the mass of the black hole, inferred either from gas kinematics or the stellar velocity dispersion (\emph{filled symbols}) or from the $K$-band luminosity of the host galaxy's bulge (\emph{empty symbols,} corrected by a factor of 0.35).  \label{F:Mbh_deltaMbh}}
\end{figure}

\begin{figure}
\plotone{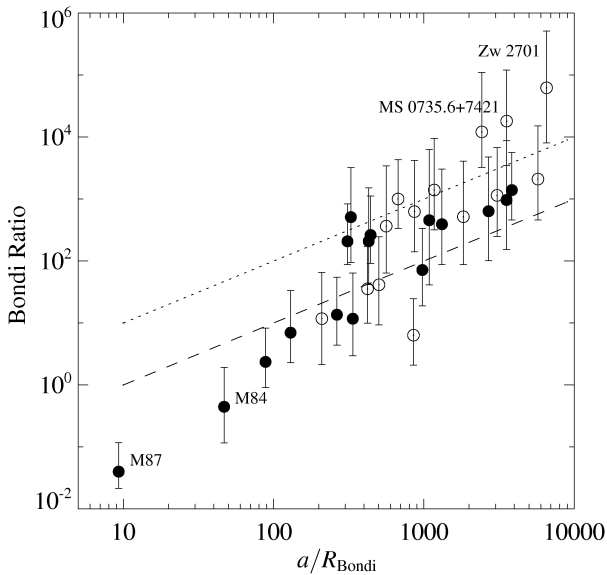}
\caption{Bondi ratio (defined as $\dot{M}_{\rm acc}/\dot{M}_{\rm Bondi}$) versus the ratio of the semi-major axis of the central region (within which the Bondi rate was calculated) to the Bondi radius. The symbols are the same as those in Figure \ref{F:Mbh_deltaMbh}. The lines denote the likely scaling of the measured Bondi ratio with the size of the central region, assuming a true Bondi ratio of 1 at the Bondi radius and a density profile $\rho \propto r^{-1},$ with either a flat core inside $a/R_{\rm{Bondi}} = 10$ (\emph{dashed line}) or no core (\emph{dotted line}). \label{F:Bondi}}
\end{figure}

\begin{figure*}
\plottwo{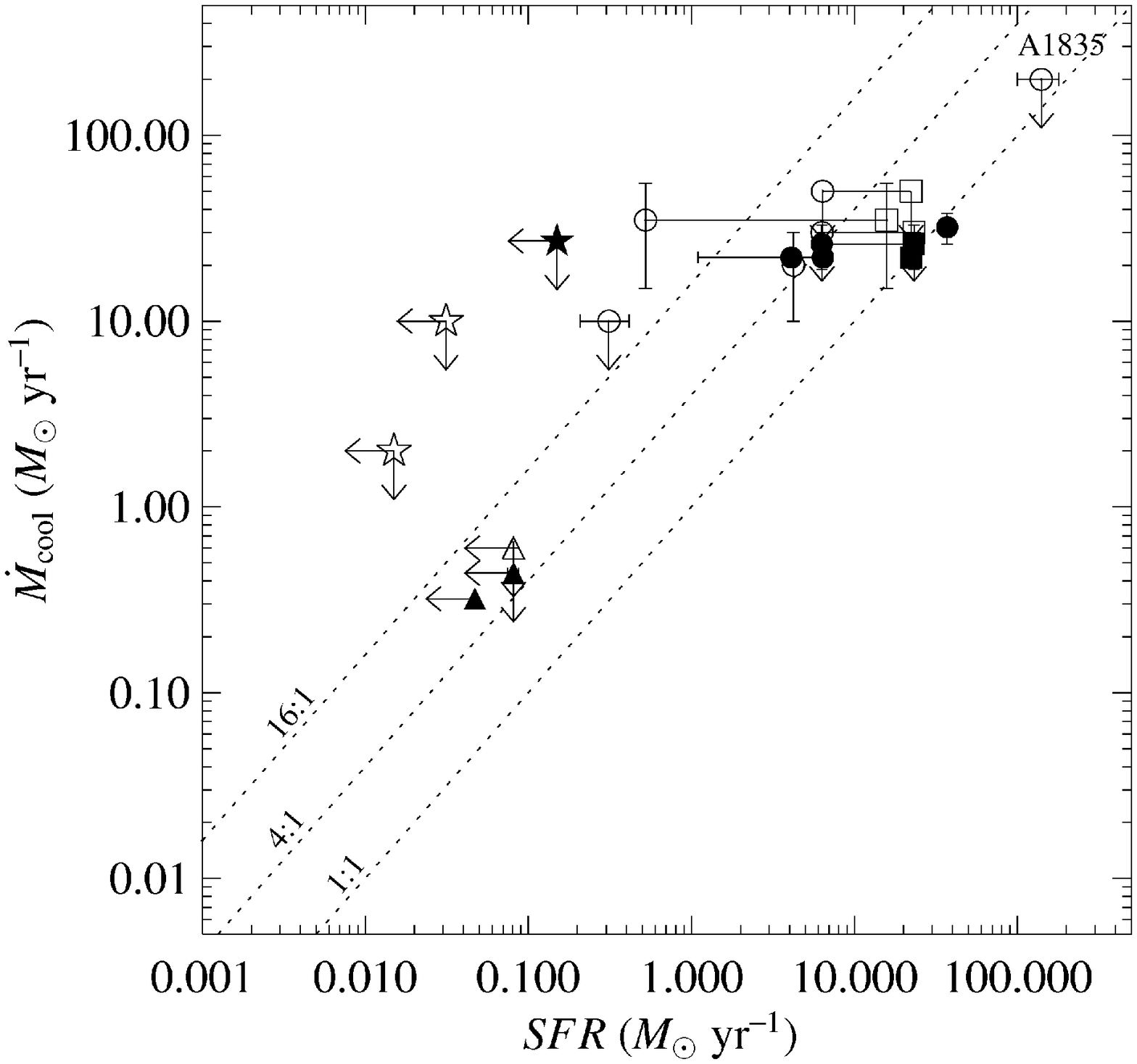}{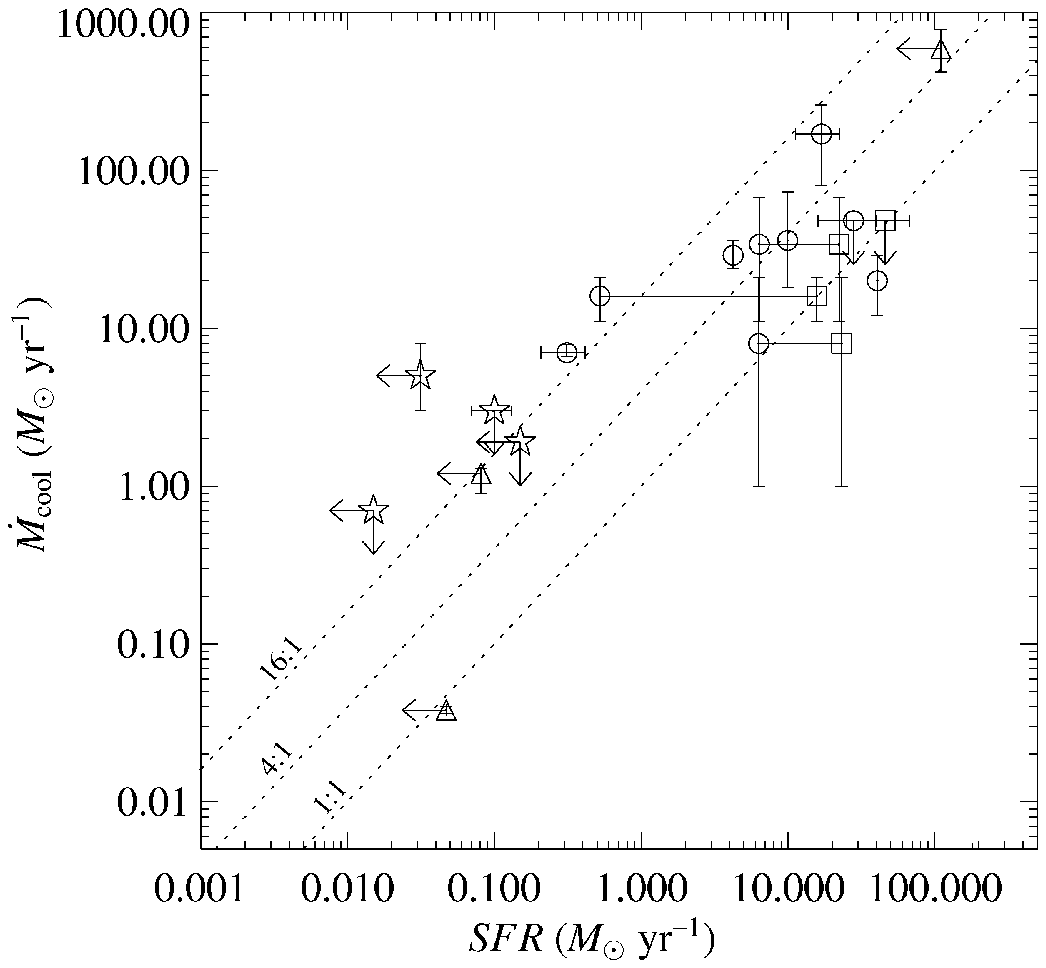}
\caption{\emph{Left:} Net cooling rate from XMM-\textit{Newton} (\emph{empty symbols}) and FUSE data (\emph{filled symbols}) versus the star formation rate. \emph{Right:} Net cooling rate from our \textit{Chandra} X-ray analysis versus the star formation rate. The symbols are the same as those in Figure \ref{F:SFR_Mdot}. The diagonal lines denote different ratios of the cooling to star formation rate. \label{F:SFR_Mdotc}}
\end{figure*}

To investigate whether the growth of the black hole, or equivalently the energy of its outburst, depends on its mass (inferred assuming bulges come equipped with mature black holes, see Section \ref{S:BH_masses}), we plot in Figure \ref{F:Mbh_deltaMbh} the fractional change in the black hole's mass during the outburst ($\Delta M_{\rm{BH}}/M_{\rm{BH}}$) against its mass.  There is no clear indication in Figure \ref{F:Mbh_deltaMbh} that the growth of the black hole depends on the black hole mass, at least to the extent that the bulge velocity dispersion or luminosity is a good black hole mass estimator for these systems. For example, systems that differ by a factor of two in inferred black hole mass, such as M84 ($M_{\rm{BH}} \sim 4 \times 10^{8}$ $M_\sun$) and MS 0735.6+7421 ($M_{\rm{BH}} \sim 7 \times 10^{8}$ $M_\sun$), differ in their fractional growth by a factor of $\sim 10^5$. However, uncertainties in the black hole and accreted masses may obscure any underlying correlation. 

For a number of objects in our sample, the implied accretion rates necessary to generate the cavities are well above our Bondi accretion rates (by factors of up to $\sim 5 \times 10^{4},$ see Table \ref{T:BH_masses} and Figure \ref{F:Bondi}). Specifically, those systems with the most powerful outbursts appear to have the largest Bondi ratios, as should be expected from the small range in black hole masses. However, as discussed in Section \ref{S:Edd_Bondi_rates}, we do not resolve the Bondi radius. Therefore, our Bondi rates are probably lower that the true values, particularly in higher redshift objects and those observations with a low number of total counts (resulting in a larger central region to obtain $\sim 3000$ counts).

To illustrate the radial dependence of this correction, we plot in Figure \ref{F:Bondi} the ratio of the accretion to Bondi rate versus the semi-major axis of the central region from which the Bondi rates were calculated, normalized to the Bondi radius. The trend in this figure supports the conclusion that the Bondi radius is not resolved. Overplotted are lines denoting the scaling of the measured Bondi ratio with radius, assuming a true Bondi ratio of unity at the Bondi radius and a density profile that rises as $\rho \propto r^{-1}$ to the Bondi radius (upper line) or flattens inside $a/R_{\rm{Bondi}}=10$ (lower line), as observed in M87 \citep{dima03}. Objects near or below these lines could reasonably have ratios of order unity or less and thus be consistent with Bondi accretion. Those significantly above the lines are likely to be accreting in excess of their Bondi rates. 

All of the objects in our sample are consistent with Bondi accretion, but only if the density continues to rise as a powerlaw to the Bondi radius. The accretion rates in those objects with the least powerful outbursts (such as M84 and M87) are generally consistent with Bondi ratios of significantly less than unity. This conclusion is supported by \citet{alle06}, who find that accretion rates in ellipticals with low-power outbursts are consistent with Bondi accretion. 

However, a number of objects (typically those with powerful outbursts, such as MS 0735.6+7421 and Zw 2701) are barely consistent with Bondi accretion and would have difficulty fueling their outbursts through Bondi accretion alone, suggesting some other route for much of the accreting material, such as cold accretion \citep[e.g., the cold feedback mechanism of][]{pizz05}. Additionally, the Bondi accretion rate assumes spherically symmetric, radial accretion, while real astrophysical flows will have some net angular momentum. An example is M87, which appears to posses a central disk of gas \citep{harm94,macc97}; thus, any accreting material would be likely to have significant angular momentum \citep[for a discussion, see][]{pizz05}. Recent hydrodynamic simulations of accretion flows \citep[e.g.,][]{prog03,krum05} find that even small amounts of angular momentum can reduce the accretion rate to well below the Bondi rate. 

It is also possible that the Bondi rates (and hence central densities) in these objects were higher at the time of the outburst than they are now. We note however that very high densities imply very short cooling times. At sufficiently high densities, the gas will cool and fall out of the hot phase in which Bondi accretion operates, placing an upper limit on the density appropriate for use in the Bondi calculation \citep[the maximal cooling flow, see][]{nuls00}. For example, to fuel the outbursts in MS 0735.6+7421 and Zw 2701 by Bondi accretion alone, the accretion rate would need to be very close to the maximal cooling flow value \citep[$\sim 10$\% of the Eddington rate for these objects, see][]{nuls00}. However, this constraint is not severe enough to rule out Bondi accretion as a viable accretion mechanism in most of our sample.

\subsection{Star Formation and Net Cooling of the ICM}\label{S:SF_Cooling}
We wish to test the hypothesis that star formation is fueled by gas condensing out of the ICM. If true, and cooling and star formation vary slowly with time, their rates should be comparable to each other. To make this comparison, we plot the net cooling (condensation) rate against the star formation rate in Figure \ref{F:SFR_Mdotc}. Symbols denote the various types of data used to derive the star formation rate (see Section \ref{S:SFR_analysis}). The cooling rates include estimates inferred from XMM-\textit{Newton} X-ray spectra and from FUSE ultraviolet spectra (see Section \ref{S:Net_cooling}), and from our \textit{Chandra} data (see Section \ref{S:Cooling_analysis}). In almost all cases, the X--ray-derived cooling rates should be considered upper limits, since indisputable evidence of cooling below $\sim 1$ keV has yet to be found in the X-ray emission from cooling flow clusters. The apparent trend should be interpreted cautiously in this context (see caveats in Section \ref{S:Net_cooling}). 

Figure \ref{F:SFR_Mdotc} shows the condensation and star formation rates for all systems in our sample are consistent with the hypothesis that star formation is fueled by gas condensing out of the ICM. The average ratio of condensation to star formation rate for those rates derived in similar apertures is $\dot{M}_{\rm{cool}}/SFR \sim 4$, using XMM-\textit{Newton} and FUSE rates (the ratio does not change significantly if \textit{Chandra} rates are considered). This value is similar to that found by \citet{hick05} in a study of star formation and cooling using XMM-\textit{Newton} UV monitor data. 

Figure \ref{F:SFR_Mdotc} shows that the rates of star formation and cooling have converged greatly and are in rough agreement in several systems. The classical cooling flow problem, in which the X--ray-derived cooling rates were factors of $10-100$ in excess of the star formation rates in most systems, has largely disappeared. While the average discrepancy of four to one is still large, it is of the order of the uncertainty in the rates. Factors that may contribute to scatter in the rates are time-dependent effects such as radio-triggered star formation \citep{mcna93} and the time lag required for gas at $\sim 10^{7}$~K to cool and form stars.

Lastly, it is clear that if star formation is being fueled by the ICM, firm detections of cooling out of the X-ray band should be within reach of present and future X-ray observatories for those objects with large star formation rates \citep{mcna06}. Even with present-day instruments, the upper limits on cooling derived to date are very close to the inferred total star formation rates for a number of objects (e.g., A1835). If this star formation scenario is to survive, future deep X-ray observations of these objects (with Constellation-X, for example) should detect this cooling gas.

\begin{figure}
\plotone{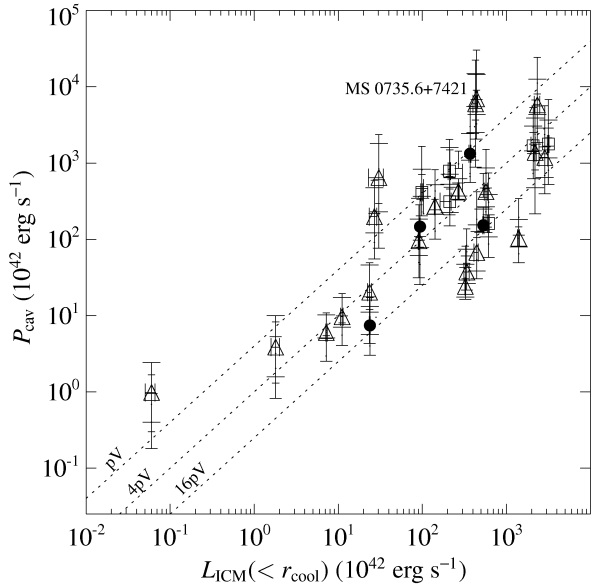}
\caption{Cavity power of the central AGN versus the X-ray luminosity of the intracluster medium inside the cooling radius that must be offset to be consistent with the spectra  ($L_{\rm{ICM}} = L_{\rm{X}} - L_{\rm{cool}}$). The symbols and wide error bars denote the values of cavity power calculated using the buoyancy timescale. The short and medium-width error bars denote the upper and lower limits of the cavity power calculated using the sound speed and refill timescales, respectively. Different symbols denote different figures of merit: \emph{circle} -- well-defined cavity with bright rims, \emph{triangle} -- well-defined cavity without bright rims, \emph{square} -- poorly defined cavity. The diagonal lines denote $P_{\rm{cav}}=L_{\rm{ICM}}$ assuming $pV,$ $4pV,$ or $16pV$ as the total enthalpy of the cavities.\\ \label{F:Lcool_Lcav}}
\end{figure}

\subsection{Quenching Cooling Flows}\label{S:quenching}
We have demonstrated that, in many systems, the net ICM cooling (condensation) rate is in rough agreement with total star formation rate. However, we have not dealt with the question of what maintains the bulk of the ICM at X-ray temperatures, preventing it from cooling out at the expected classical rates (typically $\sim 10-100$ times the star formation rates). AGN heating, through cavities, shocks, and sound waves, has emerged as the favored mechanism to prevent this massive cooling in cooling flows.
To investigate whether AGN cavities are powerful enough to balance the radiation emitted by the ICM, we plot in Figure \ref{F:Lcool_Lcav} the cavity power of the central AGN against the total radiative luminosity of the intracluster gas within the cooling radius (minus the luminosity due to net cooling, given in Table \ref{T:properties}). This plot supersedes Figure 2 of \citet{birz04}, to whose sample we have added deeper X-ray data and 14 new cavity systems, most of which lie in the upper half of cavity powers.

Remarkably, most of the systems in our sample have cavity powers sufficient or nearly sufficient to balance the entire radiative losses of the ICM within the cooling radius. The remaining systems may require other forms of heat to offset cooling completely, such as thermal conduction \citep{voig04}. However, we note that the time-dependent nature of AGN feedback does not require that cooling is always balanced by heating. It is possible that those systems that do not currently balance are in a cooling phase and will be entering a heating phase soon. Intermittent heating and cooling would allow for cooling and star formation at observed levels.

In Figure \ref{F:Lcool_Lcav}, a number of systems lie well above the $4pV$ line and even above the $pV$ line, implying that their cavities likely represent more energy than required to balance cooling. These systems, many of which possess supercavities and shocks extending beyond the cooling radius, have enough energy to quench cooling and to contribute to cluster preheating. An example is MS 0735.6+7421, the most powerful such outburst know to date. The AGN in this cluster has dumped $\sim 1/3$ keV per particle into the ICM \citep[including the energy of the shock;][]{mcna05}. The cavities alone have enough energy to quench cooling 15 times over. This amount of energy, even if distributed only partly inside the cooling radius, should have a profound effect on any cooling gas. Such objects may thus be in a heating phase.

However, unresolved problems still remain, the most obvious of which is the absence of cavities and star formation in many cooling flow systems. The absence of cavities currently does not however rule out significant feedback in the past. \citet{dona06} find elevated entropy levels throughout the cooling region of both cooling-flow clusters that show evidence of AGN feedback and those that do not, consistent with a history of AGN feedback in all cooling flows. Secondly, in cooling flows that lack evidence of AGN heating, other sources of heat, such as thermal conduction, may be important. Lastly, the most powerful explosions, such as seen in MS 0735.6+7421, may turn off accretion and hence AGN activity for extended periods.   

\section{Conclusions}
We have presented an analysis of the star formation and AGN properties in 33 CDGs in the cores of cooling flows. We find that the AGN outbursts in most of the systems have enough energy to offset most of the radiative losses of the ICM, and to severely reduce cooling to levels that are approaching the star formation rates in the central galaxy. Using the cavities to infer black hole growth and star formation to infer bulge growth, we find that bulge and black hole growth rates scale roughly with each other in rough accordance with the slope of the Magorrian relation. The large scatter may indicate that growth occurs in spurts, with periods of cooling and star formation interspaced with periods of heating, or that the efficiency of the conversion of the gravitational binding energy of the accreting matter to outburst energy varies across the sample. We find the central supermassive black holes are growing at rates of $\sim 10^{-4}$ $M_{\sun}$ yr$^{-1}$ to $\sim 1$ $M_{\sun}$ yr$^{-1}$ (with a median rate of 0.035 $M_{\sun}$ yr$^{-1}$), which, in most of our sample, are insufficient to account for their present-day masses. However, a number of black holes are growing at rates that are consistent with their formation from scratch in $\sim 10^{10}$ yr. The extreme cases are those objects experiencing the most powerful outbursts ($P_{\rm{cav}} \sim 5 \times 10^{45}$ ergs s$^{-1},$ approximately 10\% of our sample), which are growing at rates sufficient to assemble their black holes in $\sim 10^9$ yr. 

Across our sample, the inferred black hole accretion rates are well below their Eddington limits but above their Bondi rates. \textit{Chandra} does not resolve the Bondi radius in these systems, and thus significant Bondi accretion cannot be ruled out. The exceptions are those systems with powerful outbursts, where either direct accretion from the hot ICM is not the principle route of cooling gas or their central properties were very different at the time of the outburst than those of typical nearby CDGs such as M87.

We test the scenario that the active star formation is fueled by cooling (condensation) from the ICM. We find that star formation and cooling rates are converging (to an average ratio of cooling to total star formation rate of four to one), and in some cases are consistent with one another. Inhomogeneities in star formation rates and the lack of firm detections of cooling in X-ray data are the main factors that limit our conclusions. Nevertheless, this rough agreement is far different from the situation a decade ago, when the best X-ray cooling rates were tens to hundreds of times the star formation rates. 

Using the best X-ray data to date, we extend and revise the heating versus cooling plot of \citet{birz04}. Remarkably, we find that AGN heating, as traced by the power in X-ray cavities alone, is capable of balancing the radiative losses of the ICM in more than half of the systems in our sample. However, the means by which the AGN's jet energy is converted to heat in the ICM and the efficiency of this conversion are not yet clear \citep[e.g.,][]{reyn02}. Additionally, our estimate of AGN heating neglects other significant sources of heat that are likely to be present in many of the systems in our sample, such as weak shocks \citep[e.g.,][]{mcna05,nuls05a,nuls05b,form05}, sound waves \citep[e.g.,][]{fabi06}, and thermal conduction \citep[e.g.,][]{voig04}. All of these heat sources may play a role in maintaining the rough balance of heating to cooling observed to exist throughout the cooling region. AGN, however, have emerged as the most important heating mechanism in cooling flows.

A unified picture of star formation, cooling, and AGN feedback is now emerging, one with applications to the more general problems of galaxy formation and the truncation of the high end of the luminosity function of galaxies. Both simulations and models of galaxy formation \citep[e.g.,][]{balo01,sija06,voit05} conclude that AGN heating is required to prevent the overcooling problem in CDM models, in which too many large galaxies are formed. A plausible scenario is that AGN regulate the cooling of gas in the cores of cooling flows, preventing most of the ICM from cooling but allowing some net condensation that feeds both star formation and black hole growth, possibly in an intermittent manner, along the Magorrian relation.

In summary, we find that star formation and cooling rates and AGN outburst energies for our sample of CDGs are broadly consistent with the simple AGN-ICM feedback scenario in which gas cooling out of the ICM feeds AGN outbursts that heat the gas in the cluster's core. Some low-level, net cooling may still proceed, and upper limits on its rate are consistent with the scenario in which net cooling is the source of material for active star formation, unusual in most ellipticals but present in many CDGs.

\acknowledgments
We gratefully acknowledge assistance from and useful discussions with Laura B\^{\i}rzan. This work was funded by NASA Long Term Space Astrophysics Grant NAG4-11025 and \textit{Chandra} General Observer Program grants AR4-5014X and GO4-5146A.

\begin{appendix}
\section{Cavity and Central ICM Properties.}
Table \ref{T:cavities} lists the properties of each cavity measured from \textit{Chandra} images. Errors in $pV$ include an estimate of the projection effects \citep[see][]{birz04}. 

\LongTables
\def\arraystretch{1.35}
\begin{deluxetable}{lcccccccc}
\tabletypesize{\scriptsize}
\tablewidth{0pt}
\tablecaption{Cavity Properties. \label{T:cavities}}
\tablehead{
	\colhead{} & \colhead{Cavity} & \colhead{$a$\tablenotemark{b}} & \colhead{$b$\tablenotemark{c}} & \colhead{$R$\tablenotemark{d}} & \colhead{$pV$} & \colhead{$t_{\rm c_{s}}$} & \colhead{$t_{\rm refill}$} & \colhead{$t_{\rm buoy}$} \\
	\colhead{System} & \colhead{FOM\tablenotemark{a}} & \colhead{(kpc)} & \colhead{(kpc)} & \colhead{(kpc)} & \colhead{($10^{58}$ erg)} & \colhead{($10^7$ yr)} & \colhead{($10^7$ yr)} & \colhead{($10^7$ yr)} }
\startdata
         A85 & 2 &   8.9 &   6.3 &  21 & $    1.2_{-    0.4}^{+    1.2}$ &  2.3 &  5.1 &  4.2 \\
        A133 & 3 &  41 &  21 &  32 & $   24_{-    1}^{+   11}$ &  3.8 & 14 &  5.1 \\
        A262 & 2 &  5.4 &   3.4 &   8.7 & $    0.060_{-    0.017}^{+    0.050}$ &  1.5 &  2.9 &  1.7 \\
             & 2 &   5.7 &   3.4 &   8.1 & $    0.065_{-    0.016}^{+    0.048}$ &  1.4 &  2.8 &  1.6 \\
     Perseus & 1 &   9.1 &   7.3 &   9.4 & $    3.7_{-    1.7}^{+    4.7}$ &  1.0 &  4.9 &  1.6 \\
             & 1 &   8.2 &   4.7 &   6.5 & $    1.6_{-    0.1}^{+    1.0}$ &  0.7 &  3.6 &  1.1 \\
             & 2 &  17 &   7.3 &  28 & $    3.9_{-    0.1}^{+    3.5}$ &  3.1 & 10 &  8.3 \\
             & 2 &  17 &  13 &  39 & $    9.7_{-    3.4}^{+   10.4}$ &  4.0 & 13 & 10 \\
  2A 0335+096 & 2 &   9.3 &   6.5 &  23 & $    1.0_{-    0.3}^{+    1.0}$ &  3.2 &  6.3 &  5.4 \\
             & 3 &   4.8 &   2.6 &  28 & $    0.072_{-    0.002}^{+    0.037}$ &  3.7 &  4.6 & 11 \\
        A478 & 2 &   5.5 &   3.4 &   9.0 & $    0.74_{-    0.18}^{+    0.57}$ &  1.0 &  2.9 &  1.8 \\
             & 2 &   5.6 &   3.4 &   9.0 & $    0.76_{-    0.17}^{+    0.55}$ &  1.0 &  3.0 &  1.8 \\
      MS 0735.6+7421 & 2 & 110 &  87 & 160 & $ 770_{-  360}^{+ 960}$ & 13 & 58 & 26 \\
             & 2 & 130 &  89 & 180 & $ 830_{-  220}^{+ 770}$ & 15 & 66 & 33 \\
     PKS 0745-191 & 3 &  26 &  17 &  31 & $   69_{-   10}^{+   56}$ &  3.0 & 12 &  5.2 \\
     4C 55.16 & 2 &  10 &   7.5 &  16 & $    4.7_{-    1.7}^{+    4.9}$ &  1.7 &  5.6 &  3.0 \\
             & 2 &  13 &   9.4 &  22 & $    7.1_{-    2.7}^{+    7.4}$ &  2.3 &  7.3 &  4.1 \\
      Hydra A\tablenotemark{e} & 2 &  18 &  12 &  29 & $    8.1_{-    1.6}^{+    7}$ &  3.0 &  8.7 &  5.1 \\
             & 2 &  20 &  12 &  31 & $    8.6_{-    0.3}^{+    6}$ &  3.2 &  9.3 &  5.6 \\
             & 3 &  42 &  21 &  78 & $   20_{-    1}^{+    8}$ &  7.8 & 21 & 17 \\
             & 3 &  34 &  24 &  66 & $   27_{-    8}^{+   27}$ &  6.6 & 19 & 12 \\
      RBS 797 & 2 &  13 &   8.5 &  24 & $   18_{-    2}^{+   14}$ &  2.2 &  7.5 &  5.0 \\
             & 2 &   9.7 &   9.7 &  20 & $   20_{-   13}^{+   36}$ &  1.8 &  6.5 &  3.4 \\
      Zw 2701 & 2 &  46 &  41 &  54 & $  220_{-  130}^{+  340}$ &  5.2 & 23 &  7.8 \\
             & 2 &  39 &  34 &  49 & $  130_{-   70}^{+  190}$ &  4.7 & 20 &  7.2 \\
      Zw 3146 & 2 &  51 &  21 &  40 & $  170_{-   10}^{+  180}$ &  3.7 & 17 &  6.8 \\
             & 2 &  36 &  30 &  59 & $  210_{-  100}^{+  280}$ &  5.0 & 21 & 10 \\
         M84 & 2 &   1.6 &   1.6 &   2.3 & $    0.002_{-    0.0015}^{+    0.004}$ &  0.5 &  1.0 &  0.4 \\
             & 2 &   2.1 &   1.2 &   2.5 & $    0.001_{-    0.0005}^{+    0.001}$ &  0.6 &  1.0 &  0.5 \\
         M87 & 2 &   2.3 &   1.4 &   2.8 & $    0.016_{-    0.003}^{+    0.012}$ &  0.4 &  0.9 &  0.4 \\
             & 2 &   1.6 &   0.8 &   2.2 & $    0.004_{-    0.001}^{+    0.002}$ &  0.4 &  0.7 &  0.4 \\
   Centaurus & 1 &   3.3 &   2.4 &   6.0 & $    0.038_{-    0.012}^{+    0.039}$ &  1.0 &  2.2 &  1.3 \\
             & 1 &   3.3 &   1.6 &   3.5 & $    0.022_{-    0.003}^{+    0.012}$ &  0.6 &  1.5 &  0.7 \\
       HCG 62 & 2 &   5.0 &   4.3 &   8.4 & $    0.027_{-    0.015}^{+    0.039}$ &  1.8 &  2.9 &  1.5 \\
             & 2 &   4.0 &   4.0 &   8.6 & $    0.019_{-    0.013}^{+    0.034}$ &  1.8 &  2.8 &  1.6 \\
       A1795 & 3 &  19 &   7.2 &  19 & $    4.7_{-    1.6}^{+    6.6}$ &  1.9 &  6.8 &  3.7 \\
       A1835 & 3 &  16 &  12 &  23 & $   27_{-   10}^{+   30}$ &  2.1 &  8.3 &  4.1 \\
             & 3 &  14 &   9.7 &  17 & $   20_{-    6}^{+   20}$ &  1.5 &  6.5 &  2.7 \\
     PKS 1404-267 & 2 &   3.5 &   2.6 &   4.6 & $    0.054_{-    0.020}^{+    0.060}$ &  0.8 &  2.0 &  0.9 \\
             & 2 &   3.2 &   2.7 &   3.8 & $    0.062_{-    0.031}^{+    0.085}$ &  0.6 &  1.8 &  0.6 \\
 MACS J1423.8+2404 & 2 &   9.4 &   9.4 &  16 & $   15_{-    10}^{+   27}$ &  1.5 &  5.7 &  2.5 \\
             & 2 &   9.4 &   9.4 &  17 & $   14_{-    9}^{+   25}$ &  1.6 &  5.9 &  2.8 \\
       A2029 & 3 &  13 &   7.2 &  32 & $    4.8_{-    0.1}^{+    2.7}$ &  2.5 &  6.8 &  6.9 \\
       A2052 & 1 &  11 &   7.9 &  11 & $    1.2_{-    0.4}^{+    1.4}$ &  1.8 &  5.5 &  1.9 \\
             & 1 &   6.5 &   6.2 &   6.7 & $    0.53_{-    0.32}^{+    0.88}$ &  1.2 &  3.6 &  1.0 \\
       MKW 3S & 3 &  54 &  23 &  59 & $   38_{-    4}^{+   39}$ &  6.0 & 21 & 12 \\
       A2199 & 2 &  15 &  10 &  19 & $    3.7_{-    1.1}^{+    3.7}$ &  2.1 &  7.1 &  3.2 \\
             & 2 &  16 &  10 &  21 & $    3.8_{-    0.4}^{+    2.9}$ &  2.3 &  7.7 &  3.8 \\
   
   Hercules A & 3 &  26 &  21 &  60 & $   13_{-    7}^{+   18}$ &  6.1 & 18 & 13 \\
             & 3 &  47 &  19 &  58 & $   18_{-    2}^{+   22}$ &  6.0 & 19 & 13 \\
       3C 388 & 2 &  15 &  15 &  27 & $    2.9_{-    1.9}^{+    5.3}$ &  2.9 &  7.6 &  3.6 \\
             & 2 &  24 &  10 &  21 & $    2.3_{-    0.2}^{+    2.2}$ &  2.4 &  6.9 &  3.1 \\
       3C 401 & 2 &  12 &  12 &  15 & $    5.4_{-    3.5}^{+    9.8}$ &  1.6 &  6.4 &  2.1 \\
             & 2 &  12 &  12 &  15 & $    5.4_{-    3.5}^{+    9.8}$ &  1.6 &  6.4 &  2.1 \\
     Cygnus A & 1 &  29 &  17 &  43 & $   28_{-    1}^{+   18}$ &  3.4 & 15 &  8.5 \\
             & 1 &  34 &  23 &  45 & $   56_{-   13}^{+   52}$ &  3.6 & 17 &  7.8 \\
   Sersic 159/03 & 3 &  20 &  14 &  23 & $    10_{-    2}^{+    9}$ &  2.9 &  9.3 &  3.8 \\
             & 3 &  22 &  17 &  26 & $   15_{-    6}^{+   17}$ &  3.3 & 11 &  4.2 \\
       A2597 & 2 &   7.1 &   7.1 &  23 & $    1.5_{-    0.9}^{+    2.6}$ &  2.5 &  7.9 &  6.8 \\
             & 2 &  10 &   7.1 &  23 & $    2.1_{-    0.6}^{+    2.0}$ &  2.4 &  8.6 &  6.6 \\
       A4059 & 2 &  20 &  10 &  23 & $    2.2_{-    0.3}^{+    1.0}$ &  2.7 &  8.4 &  4.2 \\
             & 2 &   9.2 &   9.2 &  19 & $    0.84_{-    0.55}^{+    1.53}$ &  2.3 &  6.2 & \phn \phd \phn \phn \phd \phn \phn 3.5 
\enddata
\tablenotetext{a}{Figure of merit. The FOM gives a relative measure of the cavity's contrast to its surroundings: (1) high contrast: bright rim surrounds cavity; (2) medium contrast: bright rim partially surrounds cavity; and (3) low contrast: no rim, or faint rim surrounds cavity.}
\tablenotetext{b}{Projected semi-major axis of the cavity.}
\tablenotetext{c}{Projected semi-minor axis of the cavity.}
\tablenotetext{d}{Projected radial distance from the core to the cavity's center.}
\tablenotetext{e}{The deeper image of Wise et al.\ (2006, in preparation) of Hydra A shows two large outer cavities beyond those measured here, but for consistency we report only those cavities apparent in archival data.}
\end{deluxetable}
\\
\\
Table \ref{T:Bondi} lists the modeled temperature and density of the central region (with semi-major axis, $a$) used in the calculation of the Bondi rate. Also listed is the Bondi radius, calculated (using Equation \ref{E:Bondi_radius}) from the central temperature and the black hole mass given in Table \ref{T:BH_masses}.
\begin{deluxetable}{lcccc}
\tabletypesize{\scriptsize}
\tablewidth{0pt}
\tablecaption{Central ICM Properties. \label{T:Bondi}}
\tablehead{
	\colhead{} & \colhead{$kT$} & \colhead{$n_{\rm{e}}$} & \colhead{$a$} & \colhead{$R_{\rm{Bondi}}$}  \\
	\colhead{System} & \colhead{(keV)} & \colhead{(cm$^{-3}$)} & \colhead{(kpc)} & \colhead{(kpc)} }
\startdata
         A85 & $2.1_{-0.2}^{+0.1}$ & $0.107_{-0.008}^{+0.009}$ &   5.8 & 0.017 \\
        A133 & $1.8_{-0.1}^{+0.1}$ & $0.048_{-0.005}^{+0.004}$ &   8.0 & 0.012 \\
        A262 & $0.86_{-0.01}^{+0.01}$ & $0.065_{-0.007}^{+0.008}$ &   3.4 & 0.013 \\
     Perseus & $4.4_{-0.4}^{+0.5}$ & $0.150_{-0.005}^{+0.005}$ &  8.6 & 0.004 \\
  2A 0335+096 & $1.4_{-0.1}^{+0.1}$ & $0.056_{-0.002}^{+0.003}$ &   5.1 & 0.012 \\
        A478 & $2.7_{-0.3}^{+0.3}$ & $0.20_{-0.02}^{+0.01}$ &   5.3 & 0.010 \\
      MS 0735.6+7421 & $3.2_{-0.2}^{+0.2}$ & $0.067_{-0.003}^{+0.002}$ &  23.8 & 0.007 \\
     PKS 0745-191 & $2.6_{-0.4}^{+0.4}$ & $0.14_{-0.01}^{+0.01}$ &  11.2 & 0.013 \\
      Hydra A & $2.6_{-0.5}^{+0.8}$ & $0.15_{-0.02}^{+0.01}$ &   4.7 & 0.011 \\
      Zw 2701 & $3.3_{-0.3}^{+0.3}$ & $0.024_{-0.002}^{+0.002}$ &  37.6 & 0.006 \\
      Zw 3146 & $3.1_{-0.2}^{+0.3}$ & $0.177_{-0.007}^{+0.007}$ &  15.0 & 0.027 \\
         M84 & $0.57_{-0.01}^{+0.01}$ & $0.105_{-0.007}^{+0.007}$ &   0.9 & 0.020 \\
         M87 & $0.94_{-0.02}^{+0.02}$ & $0.191_{-0.009}^{+0.009}$ &   1.0 & 0.110 \\
   Centaurus & $0.77_{-0.01}^{+0.01}$ & $0.23_{-0.01}^{+0.01}$ &   1.3 & 0.015 \\
       HCG 62 & $0.67_{-0.01}^{+0.01}$ & $0.057_{-0.005}^{+0.007}$ &   2.1 & 0.010 \\
       A1795 & $2.7_{-0.4}^{+0.6}$ & $0.067_{-0.005}^{+0.005}$ &   9.5 & 0.007 \\
       A1835 & $4.0_{-0.3}^{+0.3}$ & $0.110_{-0.003}^{+0.003}$ &  27.2 & 0.015 \\
     PKS 1404-267 & $1.3_{-0.1}^{+0.1}$ & $0.046_{-0.002}^{+0.002}$ &   8.5 & 0.009 \\
       A2029 & $2.9_{-0.2}^{+0.3}$ & $0.37_{-0.03}^{+0.04}$ &   2.2 & 0.022 \\
       A2052 & $0.71_{-0.08}^{+0.04}$ & $0.017_{-0.002}^{+0.002}$ &   5.5 & 0.017 \\
       MKW 3S & $2.8_{-0.5}^{+0.8}$ & $0.028_{-0.009}^{+0.006}$ &  7.8 & 0.003 \\
       A2199 & $2.2_{-0.1}^{+0.2}$ & $0.099_{-0.005}^{+0.005}$ &  4.4 & 0.010 \\
   Hercules A & $2.0_{-0.2}^{+0.2}$ & $0.0111_{-0.0005}^{+0.0006}$ &  67.0 & 0.012 \\
       3C 388 & $3.0_{-0.2}^{+0.2}$ & $0.0069_{-0.0004}^{+0.0004}$ &  55.6 & 0.016 \\
     Cygnus A & $5.2_{-0.6}^{+0.5}$ & $0.132_{-0.008}^{+0.009}$ &   5.3 & 0.017 \\
   Sersic 159/03 & $1.8_{-0.1}^{+0.2}$ & $0.056_{-0.004}^{+0.004}$ &  12.2 & 0.010 \\
       A2597 & $1.6_{-0.2}^{+0.2}$ & $0.073_{-0.005}^{+0.005}$ &  11.0 & 0.006 \\
       A4059 & $2.1_{-0.1}^{+0.1}$ & $0.022_{-0.001}^{+0.001}$ &  10.6 & 0.010 
\enddata
\end{deluxetable}
\end{appendix}


\end{document}